\documentclass[aps,prb,twocolumn,superscriptaddress,longbibliography]{revtex4-2}

\usepackage{mhchem,graphicx,longtable,xfrac}
\usepackage[colorlinks=true,citecolor=blue,linkcolor=blue,breaklinks=true]{hyperref}
\usepackage{amssymb}
\usepackage{graphicx}
\usepackage{amsmath}
\usepackage{multirow}
\usepackage{tabularx}
\usepackage{array}

\usepackage{times}
\usepackage{color}
\usepackage{bm}

\begin{document}

\newcommand {\beq} {\begin{equation}}
\newcommand {\eeq} {\end{equation}}
\newcommand {\bqa} {\begin{eqnarray}}
\newcommand {\eqa} {\end{eqnarray}}
\newcommand {\ba} {\ensuremath{b^\dagger}}
\newcommand {\Ma} {\ensuremath{M^\dagger}}
\newcommand {\psia} {\ensuremath{\psi^\dagger}}
\newcommand {\psita} {\ensuremath{\tilde{\psi}^\dagger}}
\newcommand{\lp} {\ensuremath{{\lambda '}}}
\newcommand{\A} {\ensuremath{{\bf A}}}
\newcommand{\Q} {\ensuremath{{\bf Q}}}
\newcommand{\kk} {\ensuremath{{\bf k}}}
\newcommand{\qq} {\ensuremath{{\bf q}}}
\newcommand{\kp} {\ensuremath{{\bf k'}}}
\newcommand{\rr} {\ensuremath{{\bf r}}}
\newcommand{\rp} {\ensuremath{{\bf r'}}}
\newcommand {\ep} {\ensuremath{\epsilon}}
\newcommand{\nbr} {\ensuremath{\langle ij \rangle}}
\newcommand {\no} {\nonumber}
\newcommand{\up} {\ensuremath{\uparrow}}
\newcommand{\dn} {\ensuremath{\downarrow}}
\newcommand{\rcol} {\textcolor{red}}


\title{Interwoven atypical quantum states in CeLiBi$_2$}

\author{Mitchell M. Bordelon}
\affiliation{Materials Physics and Applications $-$ Quantum, Los Alamos National Laboratory, Los Alamos, New Mexico  87545, USA}

\author{Cl\'{e}ment Girod}
\affiliation{Materials Physics and Applications $-$ Quantum, Los Alamos National Laboratory, Los Alamos, New Mexico  87545, USA}

\author{Filip Ronning}
\affiliation{Materials Physics and Applications $-$ Quantum, Los Alamos National Laboratory, Los Alamos, New Mexico  87545, USA}

\author{ Km Rubi}
\affiliation{National High Magnetic Field Laboratory, Los Alamos National Laboratory, Los Alamos, New Mexico 87545, USA}

\author{Neil Harrison}
\affiliation{National High Magnetic Field Laboratory, Los Alamos National Laboratory, Los Alamos, New Mexico 87545, USA}

\author{Joe D. Thompson}
\affiliation{Materials Physics and Applications $-$ Quantum, Los Alamos National Laboratory, Los Alamos, New Mexico 87545, USA}

\author{Clarina dela Cruz}
\affiliation{Neutron Scattering Division, Oak Ridge National Laboratory, Oak Ridge, Tennessee 37831, USA}

\author{Sean M. Thomas}
\affiliation{Materials Physics and Applications $-$ Quantum, Los Alamos National Laboratory, Los Alamos, New Mexico  87545, USA}

\author{Eric D. Bauer}
\affiliation{Materials Physics and Applications $-$ Quantum, Los Alamos National Laboratory, Los Alamos, New Mexico  87545, USA}

\author{Priscila F.S. Rosa}
\affiliation{Materials Physics and Applications $-$ Quantum, Los Alamos National Laboratory, Los Alamos, New Mexico  87545, USA}

\date{\today}

\begin{abstract}
We report the discovery of CeLiBi$_2$, the first example of a material in the tetragonal Ce$TX_2$ ($T$ = transition metal; $X$ = pnictogen) family wherein an alkali cation replaces the typical transition metal. Magnetic susceptibility and neutron powder diffraction measurements are consistent with a crystal-field $\Gamma_6$ ground state Kramers doublet that orders antiferromagnetically below $T_N = 3.4$ K with an incommensurate propagation wave vector ${\bf{k}} = (0, 0.0724(4), 0.5)$ that generates a nanometric modulation of the magnetic structure. The best model of the ordered state is an elliptical cycloid with Ce moments primarily residing in the $ab$ plane. This is highly unusual, as all other $\Gamma_6$ Ce$TX_2$ members order ferromagnetically. Further, we observe an atypical hard-axis metamagnetic transition at $2$ T in magnetostriction, magnetization, and resistivity measurements. CeLiBi$_2$ is a rare example of a highly conductive material with dominant skew scattering leading to a large anomalous Hall effect. Quantum oscillations with five frequencies arise in magnetostriction and magnetic susceptibility data to $T = 30$ K and $\mu_0H = 55$ T, which indicate small Fermi pockets of light carriers with effective masses as low as 0.07$m_e$. Density functional theory calculations indicate that square-net Dirac-like Bi$-p$ bands are responsible for these ultralight carriers. Together, our results show that CeLiBi$_2$ enables multiple atypical magnetic and electronic properties in a single clean material.
\end{abstract}
\pacs{}
\maketitle

\section{Introduction}

Rare-earth intermetallics show complex quantum behavior derived from the competition of crystalline electric field (CEF) effects, Ruderman-Kittel-Kasuya-Yosida (RKKY) exchange, and Kondo interactions \cite{ruderman1954indirect, kasuya1956theory, yosida1957magnetic, mattis1962role, brando2016metallic, stewart1984heavy, coleman2012dimensions, hoshino2013itinerant, fisk1986heavy, fisk1988heavy}. The CEF electric potential splits the strong spin-orbit coupled $f$ electron total angular momentum states, which defines the local moment size and anisotropy across different local point group symmetries. In $f$ electron metals, this CEF interaction is usually on the energy scale of 1$-$100 meV \cite{cox1998exotic, han1997ab, hutchings1964point, hutchings1963investigation, rosa2015role}, which implies that the thermal population of CEF states is often relevant in determining magnetic properties. Importantly, the nature of the local moment plays a key role in setting the basis for RKKY and Kondo interactions \cite{ruderman1954indirect, kasuya1956theory, yosida1957magnetic, mattis1962role, brando2016metallic, stewart1984heavy, coleman2012dimensions, hoshino2013itinerant, fisk1986heavy, fisk1988heavy}. These competing interactions can be further tuned with external parameters, and the fine balance between these individually complex interactions leads to emergent phenomena including superconductivity \cite{gegenwart2007multiple, keimer2017physics, muller1971kondo}, quantum criticality \cite{custers2003break, si1999quantum, si2014kondo, watanabe2007fermi}, nematicity \cite{fradkin2009nematic, seo2020nematic, ronning2017electronic}, non-Fermi liquid behavior \cite{gegenwart2007multiple, paglione2007incoherent}, and incommensurate magnetic states \cite{gignoux2001frustration, donni1996geometrically, bao2000incommensurate}. 

The family of tetragonal intermetallic compounds Ce$TX_2$ ($T$ = transition metal; $X$ = pnictogen) exemplifies this interconnection with magnetic and electronic properties that are tunable in pressure and magnetic field \cite{rosa2015role, seo2020nematic, balicas2005magnetic, zhao2016field, seo2012pressure, marcus2018multi, waite2022spin, takeuchi2003anisotropic, nikitin2021magnetic, prozorov2022topological, jang2019magnetic, thamizhavel2003anisotropic, andre2000magnetic, thamizhavel2003low, piva2020electronic, adriano2015magnetic, piva2018high, adriano2014physical, freitas2020tuning, skolozdra1994cecusb2, datta2022giant, thomas2016hall, seo2021spin}. In particular, this family shows an experimental link between the $f$ electron CEF ground state and the long-range magnetically ordered states \cite{rosa2015role, freitas2020tuning}. The CEF of the Ce$^{3+}$ ions ($J = 5/2$) are split within a $C_{4v}$ point group into three Kramers doublets, wherein the ground state doublet can be either $\Gamma_6$ (pure $m_j = \pm |1/2\rangle$) or $\Gamma_7$ (mixed $m_j = \pm |5/2\rangle$ and $\pm |3/2\rangle$). Typically, $\Gamma_6$ materials are ferromagnets with Ce moments in the $ab$ plane, whereas $\Gamma_7$ materials order antiferromagnetically with moments parallel to the $c$ axis \cite{rosa2015role}. Notably, $\Gamma_6$ members have a larger $c/a$ ratio compared to their $\Gamma_7$ counterparts. 

Electrical and structural properties of the Ce$TX_2$ materials are intertwined with the CEF ground state and its corresponding magnetically ordered state \cite{seo2021spin}. For instance, in CeAgBi$_2$ ($\Gamma_7$), antiferromagnetic (AFM) order at $T_N = 6.4$~K gives rise to successive metamagnetic transitions that produce a large anomalous Hall component in an applied magnetic field \cite{thomas2016hall}. In CeAuSb$_2$, an AFM stripe nematic state that breaks fourfold rotational symmetry sets in at zero field, whereas a double-Q magnetic structure emerges at high fields \cite{seo2020nematic, balicas2005magnetic, zhao2016field, seo2012pressure, marcus2018multi, waite2022spin}. 

Studying the multitude of electronic, structural, and magnetic states in Ce$TX_2$ materials and their relation to the CEF-magnetism trend is paramount to understanding and predicting their physical properties. For instance, superconductivity has yet to be shown in the $P4/nmm$ Ce$TX_2$ series despite their apparent corollaries to $f$- and $d$-based superconductors that crystallize in the same space group \cite{kamihara2008iron, sadovskii2008high, wen2011materials, khim2021field}. Tuning quantum materials toward superconductivity or other emergent states will require a thorough description of the CEF relation not just to magnetic order but also to electronic properties. 

In particular, it is key to study Ce$TX_2$ materials wherein the CEF$-$magnetic order trend breaks down, as this is where nearly degenerate magnetic and electronic ground states may reside and generate unconventional states. One notable example is CeAgSb$_2$ ($\Gamma_6$ ground state) \cite{jang2019magnetic, takeuchi2003anisotropic, nikitin2021magnetic, prozorov2022topological}, which contains a spontaneously ordered hard-axis FM moment along the $c$ axis which contradicts the general $\Gamma_6$ $-$ FM trend. However, the moment reorients to the $ab$ plane at $\mu_0H =$ 2.8 T. The unusual zero-field FM structure has been recently suggested to derive from a dominant RKKY exchange along the $c$ axis that competes with the CEF anisotropy to determine the zero field magnetic ground state \cite{nikitin2021magnetic}. Recent reports on CeAgSb$_2$ have observed tubular shaped topological magnetic domains displaying hysteretic features arising from patterning of the topological domains related to competing $ab$ plane CEF easy axes \cite{prozorov2022topological}. This suggests that tuning the local CEF in Ce$TX_2$ materials could produce competition and phase transitions between magnetically ordered ground states and electronic states. 

Here, we report on newly-synthesized CeLiBi$_2$, a member of Ce$TX_2$ family that contradicts the CEF$-$magnetic state trend in an unfamiliar manner. CeLiBi$_2$ replaces the typical transition metal with Li (Figure \ref{fig:crystalstructure}) and creates a chemical variation in the Ce CEF environment. This material undergoes AFM order at $T_N = 3.4$ K as indicated by sharp signatures in magnetic susceptibility, specific heat, thermal expansion, and electrical resistivity measurements. Surprisingly, fits to anisotropic magnetic susceptibility show that antiferromagnetic CeLiBi$_2$ contains a $\Gamma_6$ CEF Kramers ground state doublet, opposing the general Ce$TX_2$ trend for ferromagnetism tied to the $\Gamma_6$ CEF ground states. We show that the AFM order is an incommensurate cycloidal structure with propagation wave vector ${\bf{k}} = (0, 0.0724(4), 0.5)$ originating from competing exchange interactions. We further reveal a field-induced transition along the hard $c$ axis, a large anomalous Hall effect, and quantum oscillations from ultralight pockets stemming from square-net Dirac-like Bi$-p$ bands. This material demonstrates how nonmagnetic tuning of the local Ce environment results in a competing balance of CEF anisotropy and magnetic order tied to uncommon electronic properties.

\section{Methods}

\begin{table}[b]
	\caption{Structural refinement parameters at room temperature collected on single crystals of CeLiBi$_2$ with x-ray diffraction ($R_1 = 0.0359, wR_2 = 0.0882$). The structure was refined in the $P4/nmm$ space group setting 2, and the occupancy of Li was held constant.}
	\def\arraystretch{1.2}
	\begin{tabular}{cc|ccccc}
		\hline
		\multicolumn{2}{c|}{$T$}       & \multicolumn{5}{c}{300 K}     \\ \hline
		\multicolumn{2}{c|}{$a=b$}     & \multicolumn{5}{c}{4.4843(3) \AA}  \\
		\multicolumn{2}{c|}{$c$}       & \multicolumn{5}{c}{10.8791(12) \AA}  \\ \hline
		Atom           & Wyckoff          & x    & y    & z            & $U_{iso}$ (\AA$^{2}$)  & Occupancy     \\ \hline
		Ce             & 2c	            & 3/4    & 3/4    & 0.26455(7)            & 0.0101(2)   & 0.98(4) \\
		Li             & 2a            & 3/4    & 1/4    & 0          & 0.027(9)   & 1 \\
		Bi              & 2b            & 3/4    & 1/4    & 1/2   & 0.0104(2)  & 0.98(4)  \\
		Bi              & 2c            & 1/4    & 1/4    & 0.16489(5)   & 0.0110(2)  & 0.99(4)  \\ \hline	
	\end{tabular}
	\label{tab:tabstruct}
	
\end{table}

\subsubsection{Synthesis}
Single crystals of CeLiBi$_2$ were obtained from a Li$-$Bi flux. Ce (99.95\%, AMES), Li (99.9\%, Sigma Aldrich), and Bi (99.999\%, Alfa Aesar) pieces in a 1:3.2:6.7 ratio were loaded into an alumina crucible and sealed under vacuum in a quartz ampule. The reagents were slowly heated to 900~$^{\circ}$C in 40 h to mitigate Li loss. The melt was held at 900~$^{\circ}$C for 12 h before slowly cooling at 2~$^{\circ}$C/h to 600~$^{\circ}$C. Then, the ampule was inverted, and the flux was removed via centrifuge. The plate-like CeLiBi$_2$ crystals were 0.5-1 mm on a side and showed surface oxidation if left in air. Single crystal x-ray diffraction data were collected at room temperature on a Bruker D8 Venture diffractometer with Mo K$\alpha$ radiation. CeLiBi$_2$ produced 9360 reflections, and the data were analyzed in the APEX 3 software suite with the Full matrix least squares method \cite{apex3}. The refined single crystal x-ray diffraction lattice and atomic parameters at room temperature are displayed in Table \ref{tab:tabstruct} with $R_1 = 0.0359, wR_2 = 0.0882$. Single crystals of LaLiBi$_2$, which have previously been reported \cite{PAN20061016}, were prepared in a similar manner to CeLiBi$_2$.

\begin{figure}[t]
	\includegraphics[scale=.5]{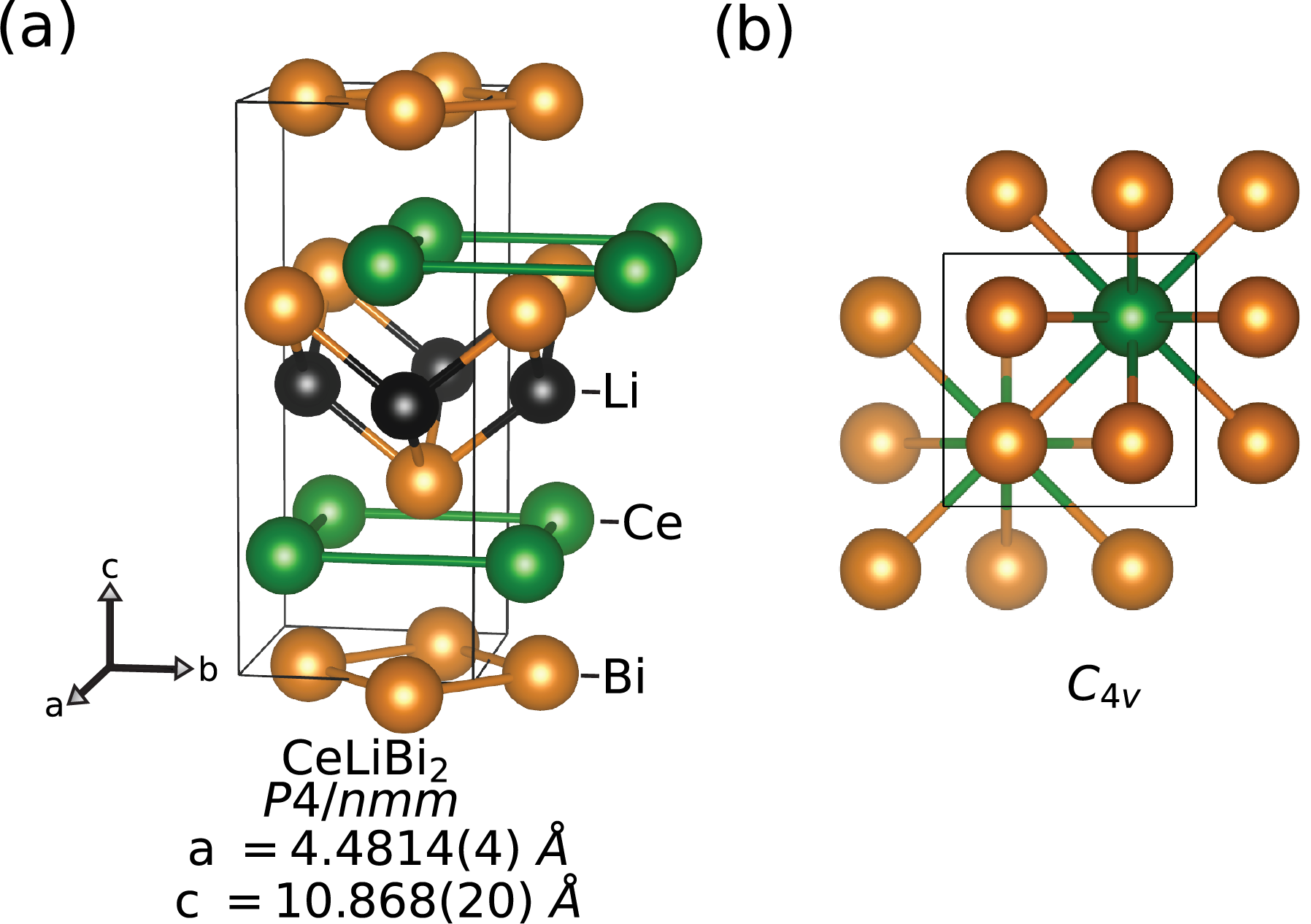}
	\caption{(a) Crystallographic structure of CeLiBi$_2$ refined from single crystal x-ray diffraction. Ce, Li, and Bi atoms are represented by green, black, and orange spheres, respectively.
		(b) View along the $c$ axis with Ce ions surrounded by eight Bi ions in a $C_{4v}$ point group. }
	\label{fig:crystalstructure}
\end{figure}

Occasionally, three other phases were observed after removing the flux, and single crystals of CeLiBi$_2$ were judiciously selected from the products. One of the phases is the known CeLi$_3$Bi$_2$ $P\bar{3}m1$ phase \cite{prakash2015synthesis}, which produced hexagonal crystals that preferentially formed when the Li content was increased and the maximum growth temperature was decreased. The two other phases are unsolved structures and were always present in a small fraction even in the optimized conditions for CeLiBi$_2$. Though the full structure of these crystals remains unknown, the lattice parameters were determined by single crystal x-ray diffraction. Flat plates crystallized in the $C2/m$ space group with lattice parameters $a = 17.377(8)$~\AA, $b = 4.536(2)$~\AA, $c = 4.714(4)$~\AA, and $\beta = 92.84(4)$$^{\circ}$. Needles crystallized in $C2/m$ with $a = 23.0279(73)$~\AA, $b = 10.2954(20)$~\AA, $c = 7.7661(20)$~\AA, and $\beta = 93.79(1)$$^{\circ}$. Both materials appear to magnetically order between 3 and 4~K. In all of our single crystal measurements, these unknown phases are not present.

Polycrystalline CeLiBi$_2$ was made from a nearly stoichiometric mixture of Ce:Li:Bi in a 1:1.2:1 ratio. The elements were loaded into an alumina crucible, sealed under vacuum, and heated to 800~$^{\circ}$C for 3 days. The furnace was then shut off and cooled to room temperature. Phase analysis was conducted on a Panalytical Empyrean powder diffractometer with Cu-K$\alpha$ radiation and analyzed with the Rietveld method in the Fullprof software suite \cite{rodriguez1993recent}.

\subsubsection{Magnetic susceptibility, heat capacity, resistivity, and dilatometry}
Magnetic properties of single crystals of CeLiBi$_2$ were collected on a Quantum Design Magnetic Properties Measurement System (MPMS3) equipped with a 7 T magnet.  Magnetic susceptibility and isothermal magnetization measurements were collected with fields parallel and perpendicular to the crystallographic $c$ axis. Isothermal magnetization data were collected at $T = 1.8, 3$ and $10$ K, and susceptibility data were collected under an applied field of $\mu_0H = $0.02~T. In poor quality samples, a superconducting transition near 2.4 K was observed, likely caused by inclusions of BiLi inside of the crystals. This transition could be suppressed under small magnetic fields less than $\mu_0H = $~0.02 T.

Specific heat measurements were collected in a Quantum Design Physical Property Measurement System (PPMS) equipped with a $^3$He insert capable of reaching 0.35 K and magnetic fields to $\mu_0H = 9$~T. Data were collected between 0.35 to 300 K. Data collected between $T = 1.8$ to $2.4$~K above $\mu_0H = 5$~T were removed due to calibration issues. For magnetic entropy analysis, the nonmagnetic analogue LaLiBi$_2$ was measured between 1.8 to 300 K. Magnetic entropy of CeLiBi$_2$ was determined by integrating $C_p/T$ after subtracting off the lattice contributions approximated by LaLiBi$_2$. 

Electrical resistivity measurements were obtained in a PPMS connected to a low-frequency AC resistance bridge. Both four-point and Hall configurations were measured with current in the $ab$ plane with external magnetic fields applied perpendicular and parallel to the $c$ axis. The crystals showed surface oxidation over time, so the surface was polished prior to attaching leads. No solvent was used to prevent material degradation.  Hall resistivity was calculated by taking the difference of positive and negative applied magnetic field data $(R_H^+ -  R_H^-) / 2$ containing the same field sweep direction.

Linear thermal expansion and magnetostriction of CeLiBi$_2$ samples were measured in a PPMS above 1.8 K using a capacitive dilatometer as described in Ref.\cite{schmiedeshoff2006versatile}. Length changes $\Delta L$ of samples with typical lengths $L =$ 300 $\mu$m and 800 $\mu$m were respectively measured along and perpendicular to the crystallographic $c$ axis. Magnetic fields up to $\mu_0H =$ 16 T were applied parallel to the measured length change. Linear thermal expansion coefficient $\alpha=\frac{1}{L}\frac{\partial \Delta L(T)}{{\partial T}}$ and linear magnetostriction coefficient $\beta=\frac{1}{L}\frac{\partial \Delta L(H)}{{\partial H}}$ were respectively computed after smoothing the field and temperature dependence of $\Delta L$.

Pulsed field de Haas-Van Alphen (dHvA) measurements were performed in a short-pulsed magnet with the sample of dimensions $\sim$0.3 $\times$ 0.3 $\times$ 0.15 mm$^3$ immersed in $^3$He liquid or gas depending on the temperature, with the temperature monitored before the pulse using a Cernox resistor. Additional dHvA measurements made in pumped $^4$He liquid confirmed that sample self-heating due to eddy currents was negligible. The detection coils of inner diameter $\sim$0.5 mm used for susceptibility measurements were wound from ASW gauge 56 with 450 turns on the inner coil and $\sim$225 turns on the outer coil to compensate for the signal originating from d$B/$d$t$.  The coils were connected to a differential amplifier with the gain set to 5000, the output of which was connected to a digitizer.

\subsubsection{Crystalline electric field analysis}

In CeLiBi$_2$, trivalent Ce ($L = 3, S = 1/2, J = 5/2$) resides in a $C_{4v}$ point group, and its CEF Hamiltonian is comprised of three parameters $B_n^m$ with Steven's operators $\hat{O}_m^n$ \cite{stevens1952matrix}:

\begin{equation}
	\label{eq:CEF}
	H_{CEF} = B_2^0 \hat{O}_2^0 + B_4^0 \hat{O}_4^0 + B_4^4 \hat{O}_4^4 
\end{equation}

\noindent This CEF Hamiltonian generates three Kramers doublets. Determination of the CEF scheme in CeLiBi$_2$ from magnetic susceptibility was conducted in the crystal field interface in Mantid Plot \cite{arnold2014mantid} combined with a numerical minimization procedure. In Mantid Plot, the magnetic susceptibility is calculated as:

\begin{equation}
	\begin{split}
		\chi (T) = \frac{N_A}{Z} \sum_i [\frac{|\langle \phi_i |g_J \mu_B \textbf{J}|\phi_i \rangle |^2}{k_B T} \\ 
		- 2 \sum_{i\ne j} \frac{|\langle \phi_i |g_J \mu_B \textbf{J}|\phi_j \rangle |^2}{E_i-E_j}] e^{- \beta E_i}
	\end{split}
\end{equation}

\noindent where $N_A$ is Avogadro's constant, $k_B$ is Boltzmann's constant, $Z$ is the partition sum, $\phi_i$ is the $i^{th}$ wave function, $E_i$ is the $i^{th}$ energy, and $\textbf{J} = J_x B_x + J_y B_y + J_z B_z$. Here, $z$ corresponds to the highest symmetry rotation axis along the $c$ axis in CeLiBi$_2$. The minimization procedure employed approaches a global minimum for the CEF parameters according to the following process adapted from Bordelon \textit{et al.} \cite{naybo2PRB}:

(1) Initialize the CEF Hamiltonian with a guess of CEF parameters, starting from values of related Ce$TX_2$ materials.

(2) Diagonalize the CEF Hamiltonian to generate CEF energy eigenvalues, wave functions, and magnetic susceptibility as a function of temperature. 

(3) Calculate $X^2 = (\chi_{calc} - \chi_{obs})^2/\chi_{calc}$ for $B \parallel c$ and $B \perp c$. This reflects the deviations between observed and CEF Hamiltonian calculated susceptibilities.

(4) Change the CEF parameters and recalculate $X^2$. If the new $X^2$ is less than the old value, accept the new CEF parameters.

(5) Iterate from multiple initialization starting points to approach a global minimum for CEF parameters.

\subsubsection{Neutron powder diffraction}
Neutron powder diffraction data were obtained on the HB2A diffractometer at the High Flux Isotope Reactor at Oak Ridge National Laboratory, TN. The $\sim$4~g CeLiBi$_2$ powder sample was loaded into a vanadium canister that was placed inside of a cryostat capable of reaching 1.5 K. Data were collected at $T = 300, 10,$ and $1.6$ K with an incident neutron wavelength of $\lambda = 2.41$ \AA \ selected by a Ge(113) monochromator. Refinement of the diffraction patterns was performed in the FullProf software suite \cite{rodriguez1993recent}. Magnetic symmetry analysis of the ordered phase was performed in the SARAh software suite \cite{wills2000new}. The structural Rietveld refinement parameters of CeLiBi$_2$ at 10 K were fixed in order to analyze the temperature-subtracted magnetic scattering at 1.64 K. 

Magnetic structure refinement excluded the extra intensity near $|Q| = 0.55$~\AA$^{-1}$ that only appears in the raw 1.64~K$-$10~K subtraction in Figure~\ref{fig:scattering}(a). It is unlikely due to magnetic ordering of CeLiBi$_2$. This intensity resides on the shoulder of the lowest angle structural peak shown in the inset of Fig. \ref{fig:scattering}(a). If the intensity was due to CeLiBi$_2$ magnetic ordering, an additional incommensurate wave vector component would be necessary to include it. However, adding intensity to the magnetic structure would cause the Ce moment size to significantly exceed the maximally allowed value from the CEF $\Gamma_6$ doublet. An estimation of the Ce moment with the included intensity can be determined by comparing the integrated intensity of the extra intensity versus the sum of the indexed CeLiBi$_2$ magnetic peaks and the extra intensity. The extra intensity accounts for roughly 40\% of the total. The expected Ce moment size would then proportionally increase by $\sim 1/0.6 = 1.67$ with a Ce moment maximal modulus of $1.67 \times 1.24 \mu_B = 2.07 \mu_B$. This far exceeds the $\Gamma_6$ CEF doublet showing that the extra peak is not intrinsic to CeLiBi$_2$ magnetic order. We conclude the extra peak is not intrinsic to CeLiBi$_2$, and two likely explanations for the extra peak are magnetism from the impurity $C2/m$ phase or structural broadening defects like stacking faults \cite{croguennec1997nature}. It was modeled at $T = 1.64$~K with increased peak broadening and asymmetry of the main structural CeLiBi$_2$ phase. The magnetic structure was fit to $1.64 - 10$~K data after subtracting the broadened peak shape model as shown in Figure~\ref{fig:scattering}(c).

\subsubsection{Density function theory}

Density functional theory (DFT) calculations were performed using the PBE exchange correlation functional\cite{PBE} as implemented in the WIEN2K software \cite{Wien2k}. Spin orbit coupling without relativistic local orbitals was included through a second variational method. To localize the $f$ electrons of Ce they were treated as core states in the calculation.

\section{Experimental Results}

\subsection{Crystal structure}
CeLiBi$_2$ crystallizes in the $P4/nmm$ tetragonal crystal structure with lattice parameters $a = 4.48$ \AA $\,$and$\,$ $c = 10.88$~\AA$\,$ shown in Figure \ref{fig:crystalstructure}. The refined single crystal x-ray diffraction lattice and atomic parameters at room temperature are displayed in Table \ref{tab:tabstruct} in the Appendix. Since Li is a light element with weak x-ray scattering, the position of the Li atom was determined by comparison with other Ce$TX_2$ materials and its occupancy was held constant during refinement. Powder neutron diffraction results detailed below refined the Li occupancy in powder samples to 0.96(6).

The CeLiBi$_2$ structure contains trivalent Ce ions in a local $C_{4v}$ point group. The nearest neighbor Ce-Ce ions reside 4.481 \AA \ apart in a square lattice, and these Ce lattices stack along the $c$ axis in a staggered fashion. Between the Ce square nets are alternating layers of LiBi and two-dimensional square lattices of Bi (Figure \ref{fig:crystalstructure}) that could form the basis for a topological Dirac semimetallic ground state \cite{tsmacs}.

\begin{figure}[]
	\includegraphics[width=\textwidth/2]{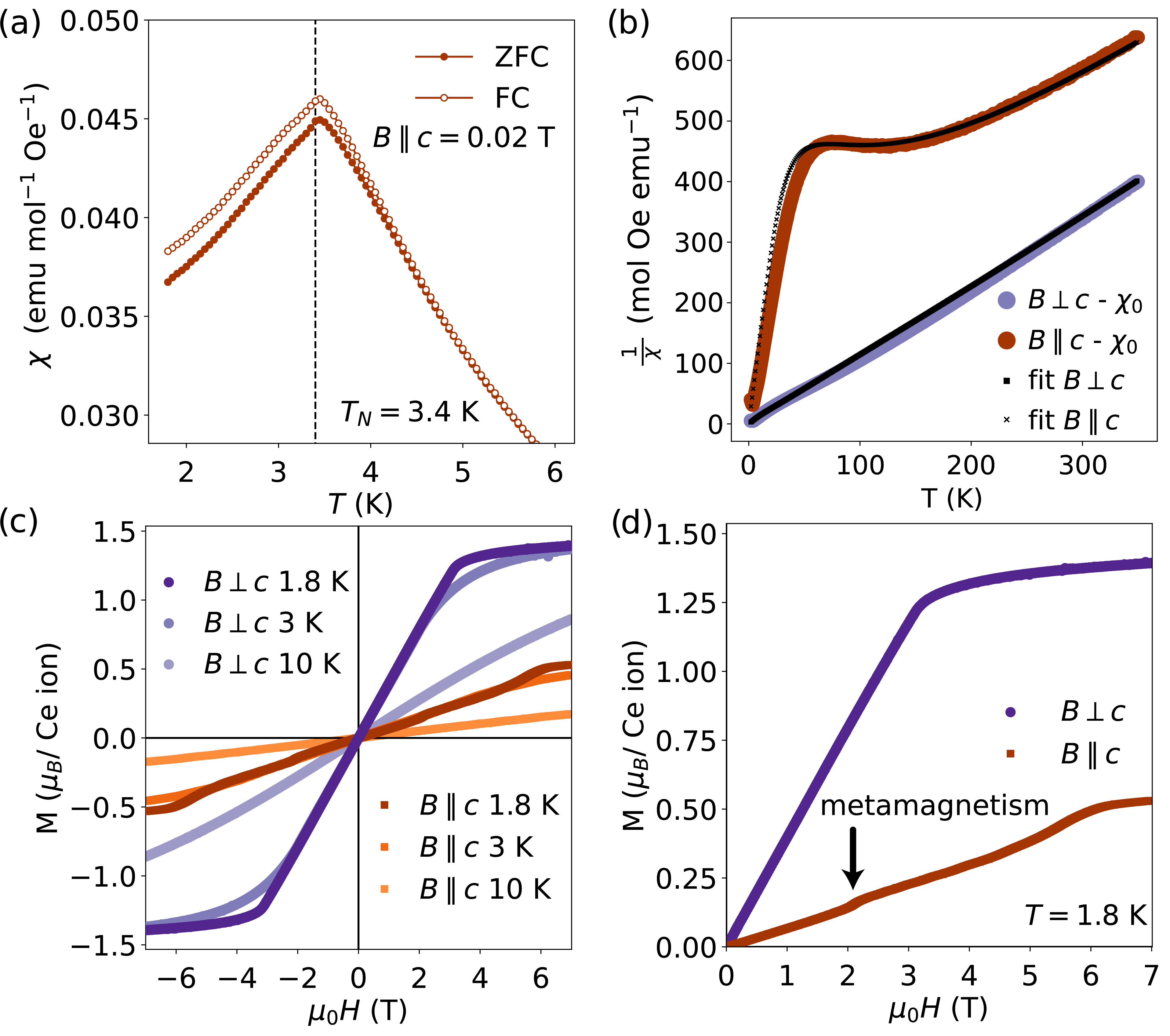}
	\caption{(a) Low-temperature magnetic susceptibility of CeLiBi$_2$ collected in a $\mu_0H = 0.02$ T field parallel to the crystallographic $c$ axis. Filled circles represent zero-field cooled (ZFC) data and open circles are field-cooled (FC) data. (b) Inverse magnetic susceptibility with fields parallel (orange) and perpendicular (purple) to the $c$ axis simultaneously fit (black) to the crystalline electric field (CEF) Hamiltonian in Eq.~(\ref{eq:CEF}). (c) Isothermal magnetization with fields parallel (orange) and perpendicular (purple) to the $c$ axis above and below $T_N$. (d) A closer look at isothermal magnetization data for $T = 1.8$ K reveals a subtle metamagnetic transition in $B \parallel c$ data at $\mu_0H = 2$ T. Above $\mu_0H_{FP} = 3$ T ($6$ T) for $B \perp c$ ($B \parallel c$), CeLiBi$_2$ enters a field polarized state.}
	\label{fig:Figure2}
\end{figure}

\subsection{Magnetic properties and crystalline electric field analysis}

The anisotropic magnetic susceptibility and isothermal magnetization data of a CeLiBi$_2$ single crystal are displayed in Figure \ref{fig:Figure2}, where $\chi_\parallel$ ($\chi_\perp$) data are collected with an applied magnetic field parallel (perpendicular) to the $c$ axis. A sharp decrease in susceptibility in Figure \ref{fig:Figure2}(a) occurs at $T_N = 3.4$ K, which suggests that CeLiBi$_2$ orders antiferromagnetically. Below $T_N$, a small splitting between zero-field cooled (ZFC) and field-cooled (FC) data arises. Figure \ref{fig:Figure2}(b) contains the magnetic susceptibility of CeLiBi$_2$ collected between 2 to 350~K. 

The powder averaged magnetic susceptibility $\chi_{pow.}~=~\frac{1}{3} \sqrt{\chi_{\parallel}^2+2~\chi_{\perp}^2}$ was fit to a Curie-Weiss law $\chi = C/(T-\theta_{CW}) + \chi_0$ in the linear region from $150-340$~K. The extracted parameters are $C = 0.750(2)$ emu K mol$^{-1}$, $\mu_{eff} = \sqrt{8C} = 2.45(1) \mu_B$, $\theta_{CW} = 5.54(32)$~K, and $\chi_0 = 0.00018(1)$ emu mol$^{-1}$. The effective moment of 2.45~$\mu_B$ is close to the expected value of $g_J \sqrt{J(J+1)} = 2.54~\mu_B$ for Ce$^{3+}$. This value is reduced likely due to the highest CEF doublet near 45.3~meV, determined from the full CEF analysis below, not being thermally occupied. The value of $\theta_{CW}$ is nonphysical, as CEF induced curvature from thermal depopulation of states offsets the Curie-Weiss law intercept.  

To fully and more precisely analyze the anisotropic magnetic susceptibility of CeLiBi$_2$, a full CEF analysis was conducted. Trivalent Ce ($J = 5/2$) resides in a $C_{4v}$ point group, and its CEF Hamiltonian in Eq.~(\ref{eq:CEF}) generates three Kramers doublets. In the $J, \ m_j$ basis, two of the doublets contain mixed $m_j = \pm |5/2\rangle$ and $m_j = \pm |3/2\rangle$ components ($\Gamma_7^1$ and $\Gamma_7^2$), and one doublet is pure $m_j = \pm |1/2\rangle$ ($\Gamma_6$). The ground state doublet selection is predominantly determined by the sign of the leading CEF Hamiltonian term, $B_2^0$. In the Ce$TX_2$ materials family, typically when $B_2^0$ is positive, the $\Gamma_6$ doublet is the ground state. If $B_2^0$ is negative, a $\Gamma_7$ ground state doublet occurs. Correspondingly, the sign of the $B_2^0$ term typically dictates the magnetic susceptibility anisotropy. When $B_2^0 < 0$, usually $\chi_\parallel > \chi_\perp$, and when $B_2^0 > 0$, usually $\chi_\parallel < \chi_\perp$. In CeLiBi$_2$, $\chi_\parallel < \chi_\perp$ with $B_2^0 > 0$, which suggests that its ground state Kramers doublet is $\Gamma_6$. 

To verify this, a fit of the CEF Hamiltonian in Eq.~(\ref{eq:CEF}) to the anisotropic magnetic susceptibility was performed with a procedure outlined in the Methods section. The resultant fit in Figure \ref{fig:Figure2}(b) has a reduced $X_{red.}^2 = X^2/\nu = 4.09$, where $\nu$ is the number of combined observable data points in magnetic susceptibility $B \parallel c$ and $B \perp c$. Extracted parameters are $B_2^0 = 1.301$~meV, $B_4^0 = -0.056$~meV, $B_4^4 = 0.728$~meV, and $\chi_0 = 0.000125$ emu mol$^{-1}$. The ground state doublet is $\Gamma_6 = |\pm 1/2\rangle$, the first excited state doublet $\Gamma_7^1 = 0.687|\pm 5/2 \rangle - 0.726|\pm 3/2 \rangle$ is at 6.1 meV, and the second excited state doublet $\Gamma_7^2 = \pm0.726|\pm 5/2 \rangle \mp0.687|\pm 3/2 \rangle$ is at 45.3 meV. 

\begin{figure}[!h]
	\includegraphics[width=\textwidth/2]{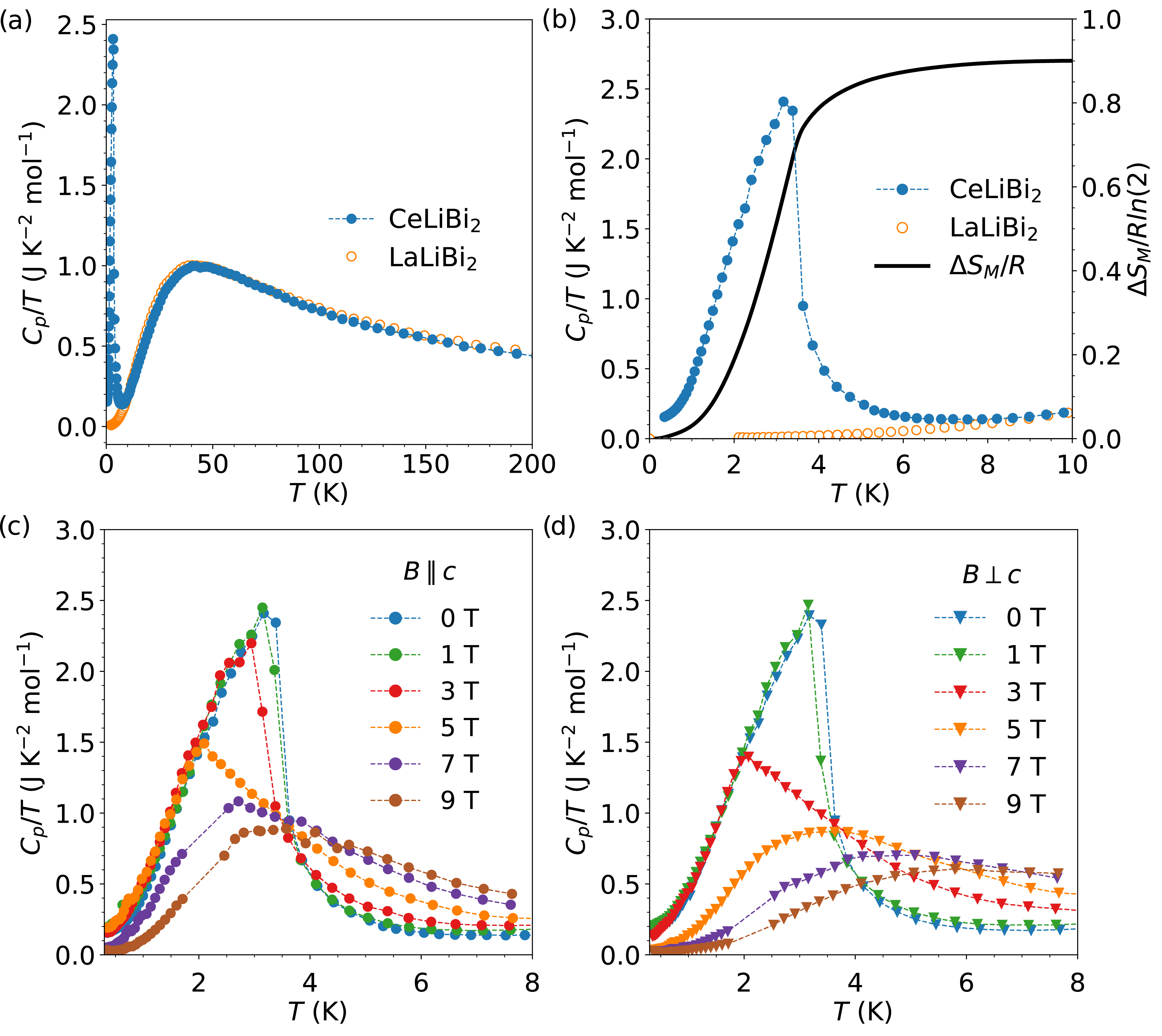}
	\caption{ (a) Temperature dependence of specific heat $C_p(T)/T$ of CeLiBi$_2$ (blue) and nonmagnetic analogue LaLiBi$_2$ (orange). (b) Between 0.35 K to 10 K, the corresponding integrated magnetic entropy ($S_M$) of CeLiBi$_2$ (black line) reaches 90\% $Rln(2)$, suggestive of weak Kondo coupling. (c)-(d) Tracking the specific heat anomaly as a function of external magnetic field applied parallel (b) and perpendicular (c) to the $c$ axis. The sharp anomaly peaked at $T_N = 3.4$~K initially shifts to lower temperature with increasing field. The sharp feature broadens out at $\mu_0H = 5$~T and $\mu_0H = 3$~T in $B \parallel c$ and $B \perp c$ data, respectively, corresponding to field polarized behavior consistent with isothermal magnetization data.}
	\label{fig:FigureCP}
\end{figure}

The projected anisotropic $g$-factor components may be calculated for the $i^{th}$ Kramers doublet as:

\begin{equation}
	\begin{split}
	g_\parallel = 2g_J| \langle \phi_i^{\pm}|J_z|\phi_i^{\pm} \rangle |, \\
	g_\perp = g_J| \langle \phi_i^{\pm}|J_{\pm}|\phi_i^{\mp} \rangle |.
	\end{split}
\end{equation}

For the $\Gamma_6$ doublet, this results with $g_{\perp} = 2.571$ and $g_{\parallel} = 0.857$. When the doublet is thermally isolated, the expected maximal ground state anisotropic magnetic moment is calculated as $m = g \mu_B J_{eff}$ where $J_{eff} = 1/2$. For the ground state $\Gamma_6$ doublet, this corresponds to $m = $1.29 $\mu_B$ perpendicular and $m = $0.43 $\mu_B$ parallel to the $c$ axis.

Isothermal magnetization M($H$) data collected above and below $T_N = 3.4$ K are displayed in Figure \ref{fig:Figure2}(c)-(d). No hysteresis is observed, which corroborates an AFM ground state. The data collected at $T = 10$ K are linear and do not saturate up to $\mu_0H = 7$~T. At $T = 1.8$ K, the data show CeLiBi$_2$ enters a field polarized (FP) regime for $B \perp c$ above $\mu_0H_{FP} = 3$~T and for $B \parallel c$ above $\mu_0H_{FP} = 6$~T. The positive non-zero slopes above these fields arise from Van Vleck susceptibility between neighboring CEF levels.  The moments reach $m_{FP} = $1.26 $\mu_B$ with $B \perp c$ and $m_{FP} = $0.39 $\mu_B$ with $B \parallel c$, indicating that the anisotropic $\Gamma_6$ Ce moments reside primarily within the $ab$ plane and is in good agreement with the CEF scheme. A closer look at the $T = 1.8$ K data in Figure \ref{fig:Figure2}(d) reveals a small hard-axis transition just above $\mu_0H = 2$~T that reappears in other thermodynamic measurements below. 

\begin{figure*}[!ht]
	\includegraphics[width=\textwidth]{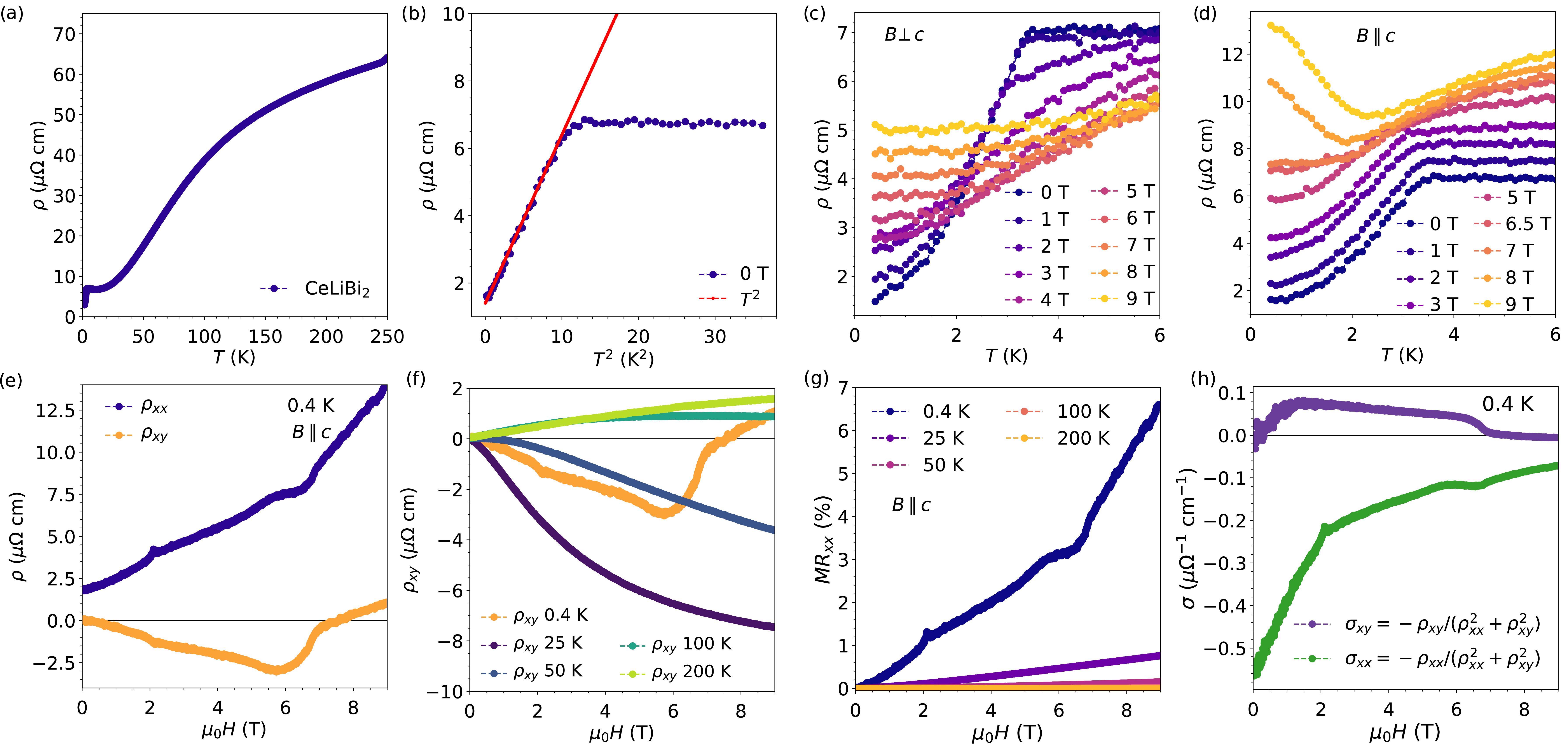}
	\caption{(a) In-plane electrical resistivity $\rho (T)$ collected on a single crystal of CeLiBi$_2$ between 2 to 250 K. Nonlinearity at high temperature ($T > 10$ K) is caused by depopulation of the two excited state CEF doublets. A sharp decrease in resistivity appears at $T_N = 3.4$ K. (b) $\rho (T)$ between 0.35 to 6 K plotted versus $T^2$ is fit to Fermi liquid behavior (red line) with $\rho_0 = 1.40$ $\mu \Omega$ cm and $A = 0.50$ $\mu \Omega$ cm K$^{-2}$. Deviations from Fermi liquid behavior due to magnetic scattering may reside below the measured temperature range. (c)-(d) Resistivity as a function of increasing magnetic field perpendicular (c) and parallel (d) to the $c$ axis shows an initial downward shift of $T_N$ and an increase in in-plane resistivity below $T_N$ with increasing field. (e) Field dependence of $\rho_{xx}$ and $\rho_{xy}$ below $T_N$ in a magnetic field parallel to the $c$ axis. $\rho_{xx}$ (purple) and $\rho_{xy}$ (orange) show two main features near $\mu_0H = 2$ and 6 T. While $\rho_{xx}$ remains positive in the full field range measured, $\rho_{xy}$ changes sign above $\mu_0H = 7$ T. (f) Magnetoresistance of CeLiBi$_2$ with varying $T$. At $T = 0.4$ K, a large positive magnetoresistance is observed that reaches nearly 650\% by $\mu_0H = 9$ T. (g) Magnetoresistance (MR) obtained at $T = 0.4, 25, 50, 100,$ and $200$ K. Below $T_N =3.4$~K, CeLiBi$_2$ exhibits large MR that approaches 650\%. h) $\sigma_{xx}$ and $\sigma_{xy}$ obtained at $T = 0.4$ K. CeLiBi$_2$ is a highly conductive material with $|\sigma_{xx}| \sim 0.55 \times 10^6$~($\Omega$~cm)$^{-1}$.}
	\label{fig:FigureRT}
\end{figure*}

\subsection{Specific heat}

Low-temperature specific heat data were obtained between 0.35 to 300 K in varying magnetic fields of $\mu_0H = 0, 1, 3, 5, 7, 9$ T. Figure \ref{fig:FigureCP}(a) shows the specific heat of CeLiBi$_2$ and nonmagnetic analogue LaLiBi$_2$ to $T = 200$ K. As shown in Figure \ref{fig:FigureCP}(b), a sharp transition occurs at zero field with its peak at $T_N = 3.4$~K (midpoint: $T = 3.5$~K). The calculated magnetic entropy was determined by subtracting off the lattice contribution to specific heat approximated by the nonmagnetic analogue LaLiBi$_2$. The magnetic entropy approaches 60\% of $Rln(2)$ at $T_N$ and 90\% of $Rln(2)$ by 7 K, suggesting that the trivalent Ce moments are well localized and short range correlations develop well above $T_N$. 
The 10\% entropy unaccounted for is caused by a small Kondo screening, residual entropy below the lowest measured temperature of $T = 0.35$~K, or by errors in the estimation of the magnetic entropy. Between 7 to 15 K, the $C/T$ versus $T$ data were fit to $C/T=\gamma+\beta T^2$ to reveal $\gamma = 42.01$ mJ mol$^{-1}$ K$^{-2}$, further indicating only a small enhancement of the effective mass $m^*$ in CeLiBi$_2$ compared to other Ce$TX_2$ materials such as CeAgSb$_2$ where $\gamma$ approaches 250 mJ mol$^{-1}$ K$^{-2}$ \cite{PhysRevB.67.224419}. 

In external magnetic fields shown in Figure~\ref{fig:FigureCP}(c)-(d), the sharp magnetic transition shifts down in temperature, again suggestive of AFM order. The feature broadens above $\mu_0H = 3$~T and $\mu_0H = 6$ T for $B \perp c$ and $B \parallel c$, respectively, whereupon CeLiBi$_2$ enters into a field polarized state. The broadened feature shifts up with increasing magnetic field in this state, as is typical of fluctuations above the ordered regime in a magnetic field \cite{vollhardt1997characteristic}. With increasing magnetic field, the magnetic specific heat component diminishes at low temperatures, and the electronic specific heat component may be more accurately determined. Analyzing the linear portion of $C_p/T$ versus $T^2$ between 0.35 to 3 K at $\mu_0H = 7$ and $9$ T with $B \parallel c$ ($B \perp c$) shows decreased $\gamma$ of 18 (10) and 5 (2) mJ mol$^{-1}$ K$^{-2}$, respectively. The decreased values of $\gamma$ compared with the zero-field value in the paramagnetic region suggests that the carriers become lighter in an external field as Kondo hybridization effects decline further.

\subsection{Resistivity}

\subsubsection{Longitudinal}
In-plane resistivity data $\rho (T)$ collected on a single crystal of CeLiBi$_2$ is presented in Figure \ref{fig:FigureRT}. At high temperatures, a broad feature centered around 100~K could indicate thermal depopulation of the excited CEF level. 
At low temperatures, $\rho (T)$ decreases sharply on cooling below $T_N = 3.4$ K as CeLiBi$_2$ magnetically orders and magnetic scattering is reduced. Below $T_N$, no magnon scattering component is observed down to $T = 0.35$~K. Figure \ref{fig:FigureRT}(b) shows $\rho (T)$ follows a $T^2$ dependence. A fit to $\rho (T) = \rho_0 + AT^2$ yields a residual resistivity of $\rho_0 = 1.40$~$\mu \Omega$~cm and a Fermi-liquid coeffcient of 
$A = 0.50$~$\mu \Omega$~cm~K$^{-2}$. No deviation from $\rho (T) \propto T^2$ was observed to 0.35 K, which, combined with a very small residual resistivity, indicates Fermi-liquid behavior and minimal magnetic scattering in a clean, highly conductive material. The estimated conductance at $T = 0$~K is $\sigma_0 = 1/\rho_0 = 0.71 \times 10^6$~($\Omega$~cm)$^{-1}$, which indicates that CeLiBi$_2$ is in the high-purity regime of $\sigma > 0.5 \times 10^6$~($\Omega$~cm)$^{-1}$ dominated by skew scattering \cite{nagaosa2010anomalous}.

Additional properties of $\rho (T)$ were studied in external magnetic fields in Figure \ref{fig:FigureRT}(c)-(d) between $\mu_0H = [0, 9]$~T and $T = [0.35, 6]$~K. In general, external fields shifted the ordering transition down in temperature, which is again an indication of antiferromagnetism. Increasing magnetic field also resulted in increased resistivity that eventually leads to an upturn in resistivity below $T = 2$ K above $\mu_0H = 7$ T with $B \parallel c$ and flattening off of resistivity with $B \perp c$. This reappears as a large positive magnetoresistance (MR) in Figure \ref{fig:FigureRT}(g) that approaches 650\% at $\mu_0H = 9$~T. The origin of increased resistivity in a magnetic field will be discussed below. 

\subsubsection{Hall resistivity}

Hall resistivity $\rho_{xy}(T)$ measurements are displayed in Figure \ref{fig:FigureRT}(e)-(f). As $T$ decreases, $\rho_{xy}(T)$ of CeLiBi$_2$ changes from hole-like to electron-like character. Below $T_N = 3.4$ K, comparison of $\rho_{xy}$ and $\rho_{xx}$ reveals two similar features. First, the metamagnetic transition, originally seen in Figure~\ref{fig:Figure2}(d) at $\mu_0H = 2$ T, is observed in both $\rho_{xy}$ and $\rho_{xx}$ data. Second, another inflection arises near $\mu_0H = 5$ to 7 T, above which the sign of $\rho_{xy}$ changes. This second feature appears just below the field polarized state in $B \parallel c$ and $T = 1.8$~K magnetization data in Figure \ref{fig:Figure2}(d). This result suggests that another phase transition occurs at $T = 0.4$~K prior to entering the field polarized regime as was similarly seen in CeAgBi$_2$ \cite{thomas2016hall}. These phase transitions reappear in $T = 0.4$~K MR data in Figure \ref{fig:FigureRT}(g). Calculated $\sigma_{xx}$ and $\sigma_{xy}$ are additionally shown in Figure \ref{fig:FigureRT}h) and further reveal that CeLiBi$_2$ is in the high-purity regime with $|\sigma_{xx}| \sim 0.55 \times 10^6$~($\Omega$~cm)$^{-1}$ at $\mu_0H = 0$~T and $T = 0.4$~K. These results further suggest Hall resistivity in CeLiBi$_2$ primarily originates from skew scattering \cite{nagaosa2010anomalous}.

\begin{figure}[]
	\includegraphics[width=\textwidth/2]{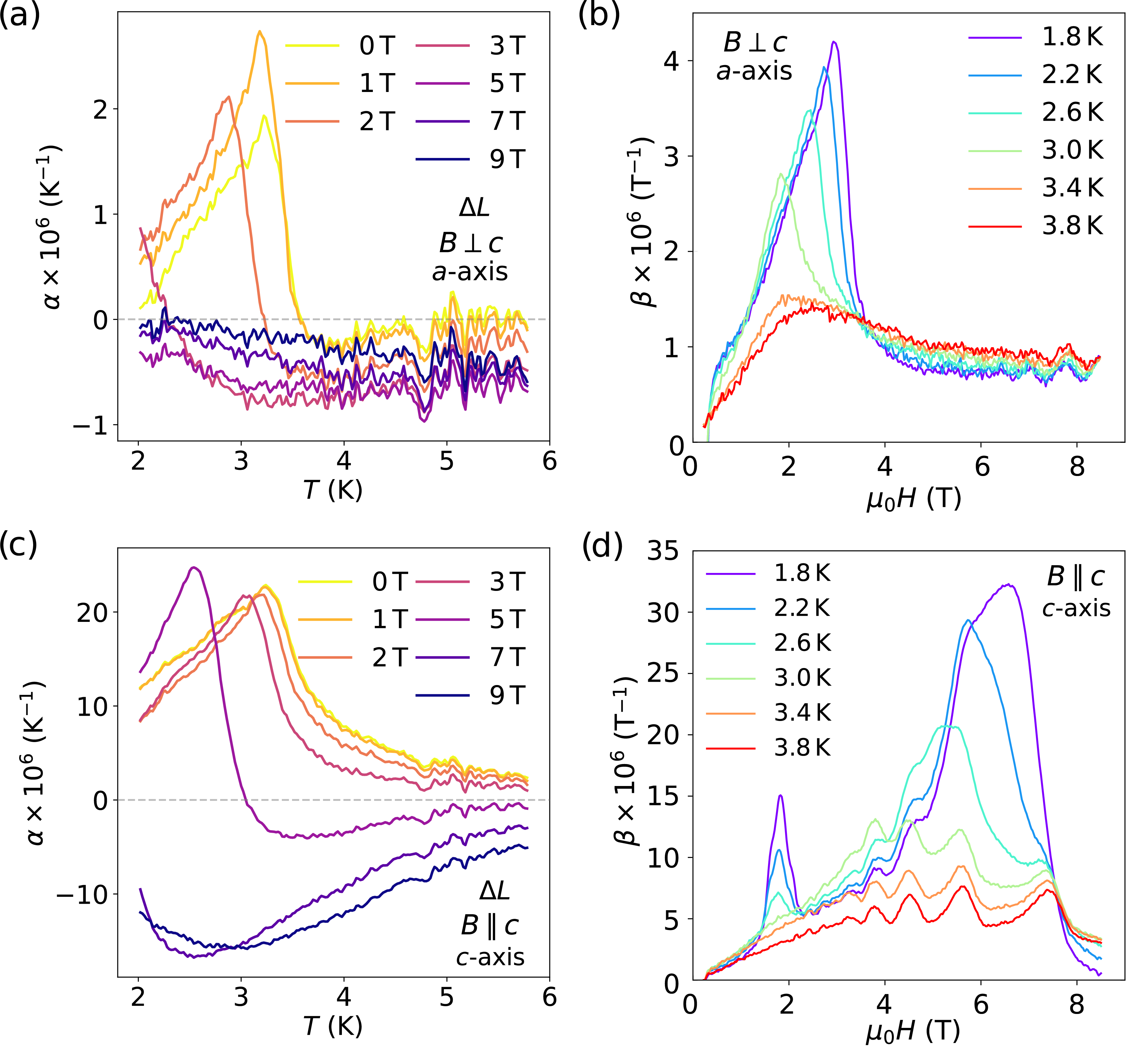}
	\caption{(a) Magnetic field dependence of linear thermal expansion coefficient $\alpha$ perpendicular to the $c$ axis tracking the evolution of $T_N$. (b) Temperature dependence of linear magnetostriction coefficient $\beta$ measured with external magnetic field and length change perpendicular to the $c$ axis. A sharp second-order phase transition appears above $\mu_0H = 3$ T at $T = 1.8$ K that shifts to lower fields and eventually broadens with increasing temperature. The transition corresponds to entering a field polarized regime as similarly observed in $B \perp c$ magnetization data in Figure 2(d). (c) Magnetic field dependence of linear thermal expansion coefficient $\alpha$ parallel to the $c$ axis with an anomaly at $T_N$. Correlated noise pattern on the thermal expansion data come from the cell effect subtraction \cite{schmiedeshoff2006versatile}, that is identical between different temperature runs. (d) Temperature dependence of linear magnetostriction coefficient $\beta$ for an external magnetic field and length change parallel to the $c$ axis. A first-order-like anomaly appears at $\mu_0H = 2$~T coincident with the $B \parallel c$ hard-axis metamagnetic transition observed in Figure 2(d). Quantum oscillations with a frequency $F \approx 23$ T are clearly visible above $\mu_0H = 3$ T and overlap with the second-order magnetic transition near $\mu_0H \approx 6-7$ T at $T = 1.8$ K that shifts to lower fields and broadens with increasing temperature.}
	\label{fig:FigureTE}
\end{figure}

\begin{figure*}[]
	\includegraphics[width=\textwidth/6*5]{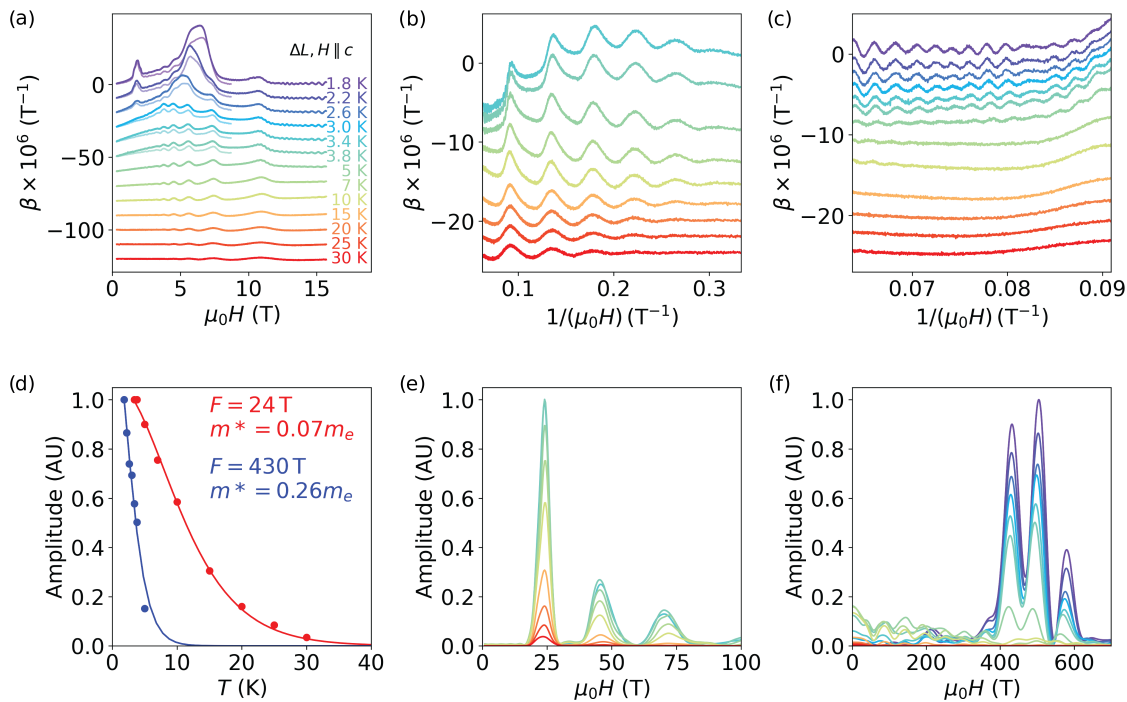}
	\caption{(a) Linear magnetostriction coefficient $\beta$ measured with field and length change along the $c$ axis at the indicated temperatures. Light colors correspond to data of Figure \ref{fig:FigureTE} (c) and dark colors to data measured on a different sample. Curves at different temperatures are offset for clarity.  (b) $\beta$ above $T = 3.4$ K as a function of $1/B$ between $\mu_0H = 3$~T  and $\mu_0H = 16$~T, displaying QO at the lowest frequency $F= 24$ T. (c) $\beta$ as a function of $1/B$ between  $\mu_0H = 11$ T  and $\mu_0H = 16$ T, displaying QO in the frequency range $F= 400-600$ T. (d) Lifshitz-Kosevich fits to the amplitudes of the $F = 24$ T (red) and $F = 430$ T (blue) QO  from panels  (b) and (c), resulting in the indicated effective masses $m^*$. (e) Frequency spectrum of $\beta(1/B)$ above $T = 3.4$ K calculated between $\mu_0H = 3$ T  and $\mu_0H = 16$ T showing the lowest frequency  $F = 24$ T and its associated harmonics at 48 and 72 T. (f) Frequency spectrum of $\beta(1/B)$ calculated between $\mu_0H = 12$ T  and $\mu_0H = 16$ T showing three nearby frequencies at $F = 430, 500$ and $580$ T.}
	\label{fig:fig_clement}
\end{figure*}

\subsection{Thermal expansion, magnetostriction, and de Haas-Van Alphen effect}

\subsubsection{Phase analysis}

Dilatometry experiments are summarized in Figure~\ref{fig:FigureTE}. Figures~\ref{fig:FigureTE}(a) and (c) show the temperature dependence of the linear thermal expansion coefficient $\alpha$ at different applied magnetic fields, with $\Delta L$ and $H$ respectively perpendicular and parallel to the $c$ axis. As for the other thermodynamic probes, we observe second-order-like anomalies at the AFM order temperature indicated by a sharp decrease in $\alpha$ near 3.4~K. The zero-field $T_N\approx 3.4$ K is in agreement with thermodynamic probes discussed above. As magnetic field increases, the ordering temperature decreases similar to the heat capacity results.

\begin{figure}[]
	\includegraphics[width=\textwidth/2]{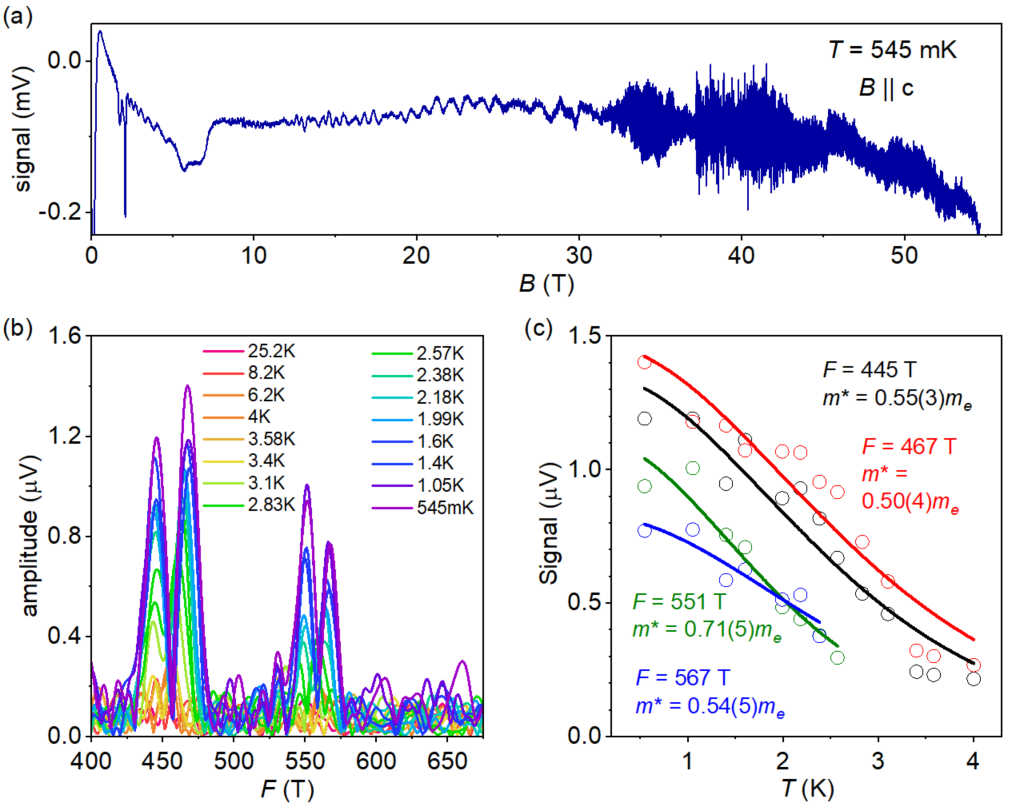}
	\caption{ (a) An example of the raw susceptibility signal measured for a sample of CeLiBi$_2$ in a pulsed magnetic field up to $\mu_0H = 55$ T and at $T = 545$ mK. (b) The cluster of $\sim 5$ dHvA frequencies obtained between 400 and 600 T obtained by taking a Fourier transform of signals versus $1/B$ for $\mu_0H = 55$ T pulses made at different temperatures as indicated. The magnetic field interval interval runs from $\mu_0H = 5$ to 45 T and the signal is modulated by a Hanning window in $1/B$. (c) The dHvA amplitude for the four dominant frequencies versus $T$ together with fits to the $T$-dependent term in the LK formula \cite{lifshitz1956theory}. The average inverse magnetic field used for the fits is $0.111$ $T^{-1}$. The fitted masses are shown together with errors. }
	\label{fig:Figure_maglab}
\end{figure}

Figure \ref{fig:FigureTE}(b) and (d) show the field dependence of the linear magnetostriction coefficient $\beta$ at different temperatures for applied field and measured length change $\Delta L$ respectively perpendicular and parallel to the $c$ axis. Typical second-order anomalies appear along both applied field directions as sharp decreases in $\beta$. This appears between 3$-$4~T for $B \perp c$ and between 6$-$8~T for $B \parallel c$ and is associated with the transition from AFM order to the field polarized regime. The value of the field where CeLiBi$_2$ enters a field polarized state obtained for $B \perp c$ at our lowest temperature $T = 1.8$ K is $\mu_0H\approx 3$ T, in agreement with the results presented above. For $B \parallel c$, clear quantum oscillations (QO) in $\beta$ are observed above $\mu_0 H = 3$ T. These oscillations, which will be discussed in following subsection, render the estimation of the onset of field polarization more difficult, especially at the base temperature $T = 1.8$ K, for which a maximum of the oscillations seats close to the magnetic anomaly. Nevertheless, the onset of field polarization can be approximated between $\mu_0H_{FP}\approx 6-7$ T that coincides with the behavior observed in $B \parallel c$ magnetization data in Figure \ref{fig:Figure2}(d). For $B\parallel c$, an additional first-order-like anomaly is found around $\mu_0H = 2$ T, arising from the hard-axis metamagnetic transition inferred from the $B \parallel c$ magnetization data in Figure \ref{fig:Figure2}(d). With increasing temperature, the onset of field polarization decreases, as expected for the weakening of the magnetic order by thermal fluctuations. 

\subsubsection{Quantum oscillations}

Extended measurements to $\mu_0H = 16$ T were carried out on a different sample to investigate the evolution of QO. A comparison between the low-field and high-field magnetostriction data of two samples is shown in Figure \ref{fig:fig_clement}(a). The almost identical amplitudes of the QO over the overlapping field range suggests that intrinsic carriers of CeLiBi$_2$, and not impurities or parasitic phases, are responsible for these QO. The low frequency QO persist at least up to $T = 30$ K. Above $\mu_0H = 10$ T, additional higher frequency oscillations are observed, but not fully resolved, to $T = 5$ K. This reveals the presence of multiple orbits with different effective masses. 
Figure \ref{fig:fig_clement}(b) shows the lowest frequency oscillations at $F= 24$ T as a function of $1/B$ between $\mu_0H = 3$~T  and $\mu_0H = 16$~T. Figure \ref{fig:fig_clement}(c) shows the highest oscillation frequencies $430-500-580$~T between $\mu_0H = 11$~T  and $\mu_0H = 16$~T.
Frequency spectra after a polynomial background subtraction contain peaks at the indicated frequencies as shown in  Figure~\ref{fig:fig_clement}(e) and Figure \ref{fig:fig_clement}(f), respectively for the low- and high-frequency ranges. The frequency of the slowest QO, $F= 24$ T, suggests a small Fermi surface on the order of $3\times 10^{-3}$ of the unreconstructed first Brillouin zone area  perpendicular to the $c$ axis. 
By fitting the temperature evolution of the amplitude of the QO displayed in Figure \ref{fig:fig_clement}(e)-(f)  to the Lifshitz-Kosevich (LK) formula \cite{lifshitz1956theory}, we obtain effective masses on the order of $m^\star \approx 0.07m_e$ for the  $F= 24$ T pocket and effective masses $m^\star \approx 0.26m_e$ for the larger $F= 430 - 500 - 580$ T pockets. 

Additional pulsed field to $\mu_0H = 55$ T de Haas-Van Alphen (dHvA) measurements are shown in Figure~\ref{fig:Figure_maglab} on another CeLiBi$_2$ sample. These measurements confirm the existence of the  $F= 24$ T QO and further reveal details of the highest QO frequencies with greater precision, indicating a total of five distinct frequencies. Quantum oscillations of $F = 23$ T begin at fields around $\mu_0H = 1$ T, but are interrupted by the magnetic phase transitions occurring at around $\mu_0H = $ 2 and $7$ T. At high magnetic fields, dHvA oscillations for a cluster of frequencies between 400 and 600 T emerge. These can be seen clearly up to $\mu_0H \approx 30$ T whereupon they become overwhelmed by much larger dHvA oscillations originating from the Cu that comprises the detection coils. The presence of frequencies between 400 and 600 T continues to be visible in Fourier transforms that are made of the data at fields extending above $\mu_0H = 30$ T. In fact, the high frequency QO is comprised of four QO frequencies split into two doublets at $F=$ 445 and 467 T and $F=$ 551 and 567~T. Their respective effective masses are shown in Figure~\ref{fig:Figure_maglab}(c) and range from $m^\star \approx 0.50 - 0.71 m_e$. These high frequency QO $m^\star$ are larger than the $m^\star \approx 0.26m_e$ mass from magnetostriction where the four frequencies are not fully resolved.

Pure bismuth is known to possess a small Fermi surface with light charge carriers \cite{edel1976electrons}. However, as stated above, QO with the same amplitudes were measured on different samples, therefore it is unlikely that these QO are coming from any type of impurity. In fact, light carriers in CeLiBi$_2$ are consistent with other thermodynamic measurements that suggest well localized Ce moments with minimal Kondo screening.

\subsection{Density functional theory analysis}

\begin{figure}[!t]
	\centering
	\includegraphics[width=3.4in]{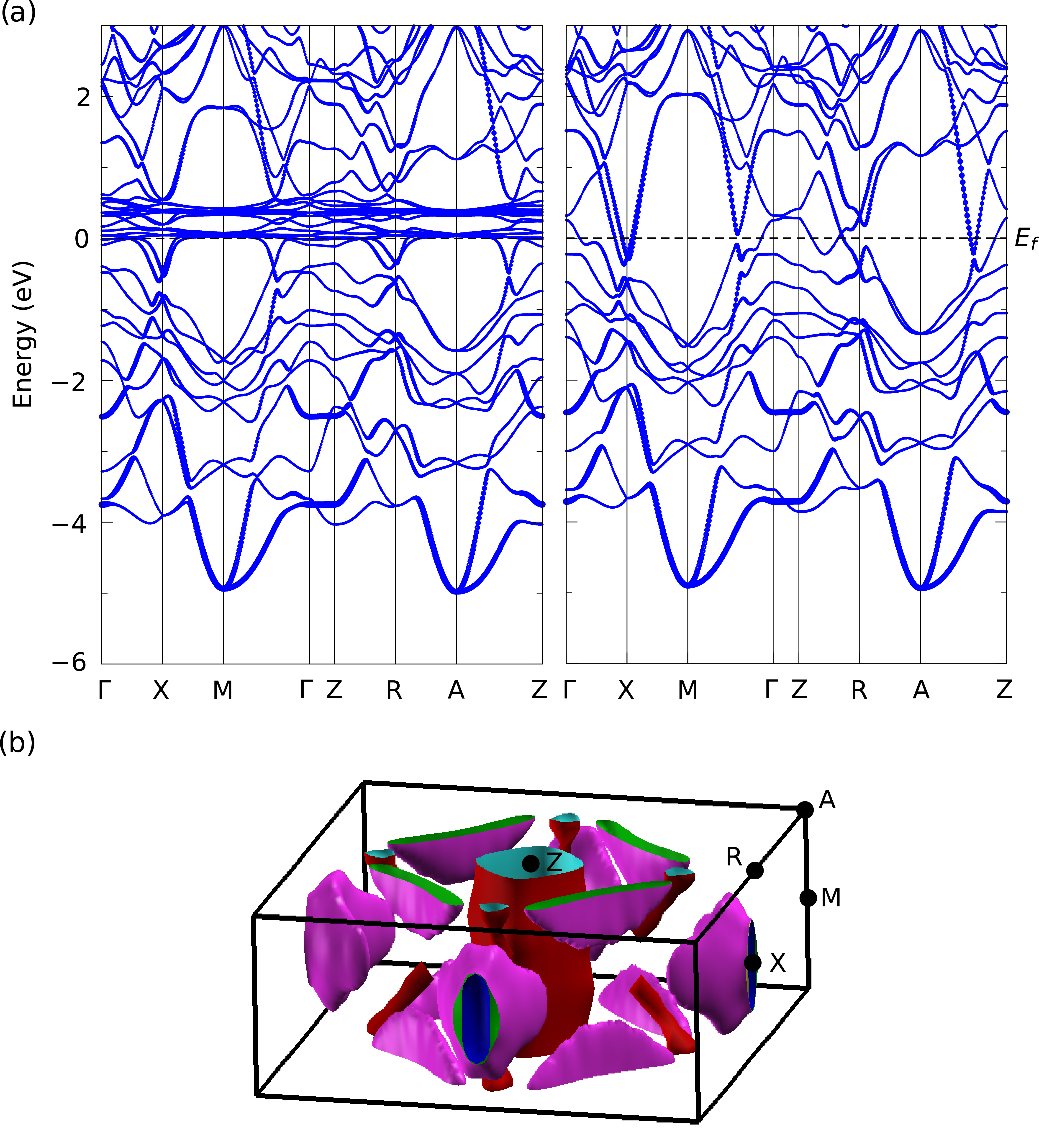}
	\caption{(a) The band structure of CeLiBi$_2$ with itinerant (left) and localized (right) $4f$ electrons where the line at zero energy indicates the Fermi level position ($E_\textrm{F}$). (b) The Fermi surface plot for the localized scenario reveals small pockets that likely generate the QO.}
	\label{fig:DFT}
\end{figure}

Figure \ref{fig:DFT}(a) shows our first-principles band structure analysis with itinerant and localized $f$ electrons. The band structure calculations include spin orbit coupling, and in general follow the previous results on structurally similar materials in (Ba,Ca,Sr)Mn(Sb,Bi)$_2$, LaAgBi$_2$, and Ce$TX_2$ materials \cite{park2011anisotropic, lee2013anisotropic, wang2012two, wang2013quasi, farhan2014aemnsb2, he2017quasi, liu2016nearly, jeong2006electronic, alsardia2020pressure}. In the absence of spin orbit coupling, linear bands cross at Dirac points along the $\Gamma - M$, $Z - R$, and $A - Z$ directions and at the $X$ point in the Brillouin zone. The primary character of the bands near the Fermi level ($E_\textrm{F}$) are from the Bi$-$5$p$ square nets. With spin orbit coupling, however, the crossings are gapped, leaving linear dispersions near $E_\textrm{F}$ but no evidence of massless Dirac fermions. The small natural electron pockets near $X$ and along $Z - R$ in Figure \ref{fig:DFT} that, depending on the tuning of $E_\textrm{F}$, are the likely source of QO data presented above.

\subsection{Elastic neutron diffraction}

\subsubsection{Crystal structure refinement}

Neutron powder diffraction results at $T = 300, 10,$ and $1.64$ K are displayed in Figure~\ref{fig:scattering}(a)-(d). Structural refinement of the 10 K data in Figure~\ref{fig:scattering}(b) required three phases, wherein CeLiBi$_2$ is the main phase ($\sim$ 81\% by mass). We emphasize that, though impurities were present in polycrystalline samples, our single crystal measurements above do not contain these phases.

The neutron powder diffraction impurity peaks were indexed to $Cmcm$ CeBi$_2$ and the lattice parameters of the needle-shaped $C2/m$ Ce$-$Li$-$Bi crystal structure (see Methods: Synthesis). No other known binary or ternary within the Ce$-$Li$-$Bi phase space could account for the observed impurity reflections. In order to model the neutron diffraction data, CeLiBi$_2$ and CeBi$_2$ were fit with the Rietveld method while $C2/m$ Ce$-$Li$-$Bi was modeled with the LeBail method \cite{le1988fourquet}. An estimate of the impurity phase fraction was determined by comparing the integrated scattering intensity corresponding to CeLiBi$_2$ and the left over intensity, resulting in roughly 81\% of the scattered intensity corresponding to CeLiBi$_2$ and 19\% to CeBi$_2$ and $C2/m$ Ce$-$Li$-$Bi. The refined lattice parameters of CeLiBi$_2$ at 10 K are $a = 4.46689(4)$ \AA \ and $c = 10.88299(15)$ \AA. Occupancies of Ce and Bi atoms refined to full within resolution of the experiment, and the occupancy of Li refined to nearly stoichiometric at 0.96(6). The overall fit had $\chi^2 = 9.63$.

Structurally-similar compounds $R$Li$_2$$Pn_2$ ($R = $ lanthanide; $Pn = $ pnictide) have been previously reported to crystallize in the $P4/nmm$ space group with the CaBe$_2$Ge$_2$ structure type \cite{fischer1982neue, ab2x2paper2}. This structure is nearly identical to that of CeLiBi$_2$ but contains a second Li site residing directly above and below the Ce ions. For CeLiBi$_2$, x-ray diffraction was insufficient to delineate these two structures due to the weak scattering intensity of light Li relative to heavy Ce and Bi. In neutron powder diffraction, Li has a significant scattering cross section and allowed for the comparison of the two structures. The Rietveld refinement of the neutron powder diffraction data in Figure \ref{fig:scattering} rules out the CaBe$_2$Ge$_2$ structure type as the second Li site occupancy refines to zero.

\subsubsection{Magnetic structure determination}

\begin{figure}[]
	\centering
	\includegraphics[width=3.4in]{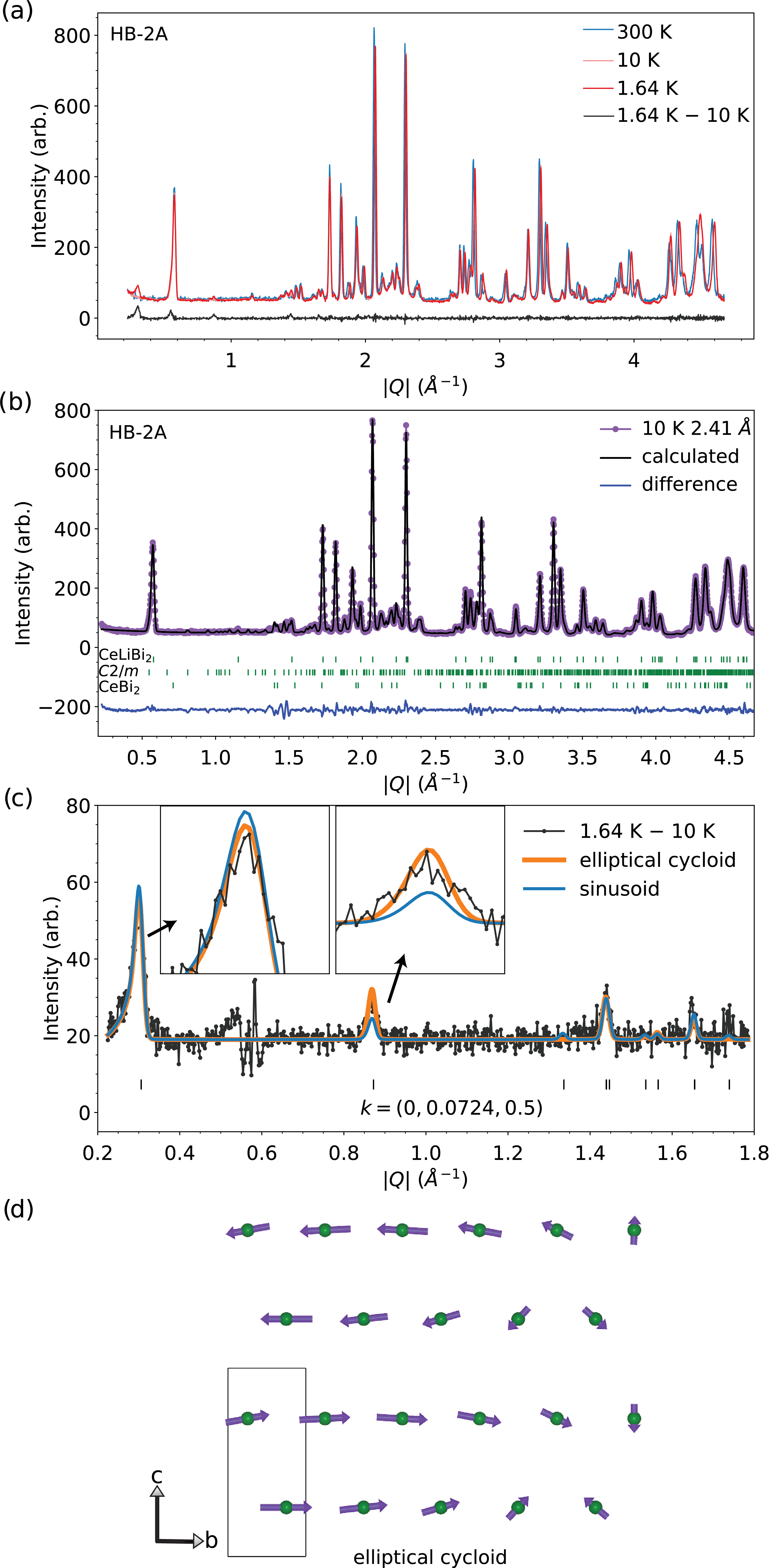}
	\caption{(a) Elastic neutron powder diffraction data collected on CeLiBi$_2$ at $T = 300, 10,$ and $1.64$~K revealing new magnetic peaks at $T = 1.64$~K. (b) Neutron powder diffraction data at $T = 10$ K. Impurity phases of CeBi$_2$ and an unsolved $C2/m$ Ce$-$Li$-$Bi compound are present, but the main phase is CeLiBi$_2$. (c) Temperature subtracted data reveal new magnetic peaks appearing below CeLiBi$_2$ $T_N = 3.4$ K fit to an incommensurate propagation wave vector ${\bf{k}} = (0, 0.0724, 0.5)$ (black ticks). These data are modeled with an elliptical cycloidal structure (orange) and a pure sinusoid structure (blue). (d) Depiction of the elliptical cycloid model where the Ce moments rotate within the $ac$ plane with the largest modulus in the $ab$ plane. The solid black lines represent the chemical unit cell. The entire magnetic unit cell doubles the chemical unit cell along the $c$ axis and is approximately 62~\AA \ long.}
	\label{fig:scattering}
\end{figure}

New magnetic peaks arise at $T = 1.6$ K as shown in Figure \ref{fig:scattering}(b). These peaks can be indexed to the antiferromagnetic incommensurate propagation wave vector ${\bf{k}} = (0, 0.0724(4), 0.5)$ except for one peak near $|Q| = 0.55$ \AA$^{-1}$, which is not intrinsic to CeLiBi$_2$ magnetic order as described in the Methods. 

With a propagation wave vector of ${\bf{k}} = (0, 0.0724(4), 0.5)$, symmetry analysis of the magnetic structure of CeLiBi$_2$ with two unique magnetic Ce atoms in SARAh \cite{wills2000new} produced four nonzero irreducible representations $IR_1$, $IR_2$,$IR_3$, and $IR_4$ (Kovalev scheme). The $IR_1$ and $IR_3$ irreducible representations contain one basis vector each while $IR_2$ and $IR_4$ contain two basis vectors each. The best fit to the data can equivalently be represented by $IR_2$ or $IR_4$, where the main difference is the stacking sequence of antiferromagnetism along the $c$ axis. $IR_2$ forms $ABBA$ coupling (AFM coupling within the atomic unit cell) while $IR_4$ forms $AABB$ coupling (AFM coupling between atomic unit cells).

Here, the fits presented in Figure~\ref{fig:scattering}(b) were produced from $IR_4$ with basis vectors shown in Table~\ref{tab:basisvectors}. Coefficients of the basis vectors $C_1$ and $C_2$ correspond to $bv^1$ and $bv^2$, respectively, and can be real or imaginary. Previous reports have shown Ce$TX_2$ materials form an elliptical cycloid \cite{unpublished} or sinusoidally modulated spin density wave \cite{marcus2018multi, waite2022spin} structure. A modulation of the Ce moment within the $IR_4$ basis can be formed by a mixture of real $C_1$ and $C_2$ of varying magnitudes (fixed moment orientation) or real $C_1$ and imaginary $C_2$ of varying magnitudes (elliptical modulation of a cycloidal structure). For reference, an equal moment cycloid forms when the magnitude of $|C_1| = |C_2|$ while $C_1$ is real and $C_2$ is imaginary.

Figure \ref{fig:scattering}(c)-(d) and Table~\ref{tab:basisvectors} show the depiction of the best fit to the data with an antiferromagnetic elliptical cycloid magnetic structure. The wavelength of the modulated cycloid along the $b$ axis is $1/0.0724 \approx 13.8$ atomic unit cells or $\lambda = b/0.0724 \approx 62$ \AA. The elliptical cycloidal model refines  $C_1 = 0.621(9)$ and $C_2 = 0.290(25)$ with a $\chi^2 = 1.99$. This model produces a Ce moment with its maximum modulus nearly parallel to the $ab$ plane ($1.24(6)$ $\mu_B$) and minimum modulus nearly parallel to the $c$ axis ($0.58(3)$ $\mu_B$), which are close to the expected anisotropic CEF values. For comparison, a pure sinusoidal model is also fit to the data in Figure \ref{fig:scattering}(c) with $C_1 = 0.539(10)$, $C_2 = 0.353(43)$, producing a marginally lower fit quality relative to the elliptical cycloid model with $\chi^2 = 2.06$.

\newcolumntype{C}{>{\raggedright\arraybackslash}m{3cm}}
\newcolumntype{G}{>{\centering\arraybackslash}m{2.5cm}}
\begin{table}[!t]
	\caption{Refined coefficients of the magnetic basis vectors of $IR_4$ creating the ellipical cycloid magnetic structure in CeLiBi$_2$ below $T_N = 3.4$~K.}
	\def\arraystretch{1.3}
	\begin{tabular}{CGG}
		\hline 
		\hline
		\multicolumn{3}{c}{$T = 1.64$~K} \\
		\multicolumn{3}{c}{${\bf{k}} = (0, 0.0724(4), 0.5)$} \\
		\hline
		\multirow{2}{*}{atom (x, y, z)}            & $bv^1$  & $bv^2$   \\
		& $C_1 = 0.621(9)$                & $C_2 = 0.290(25)$              \\
		\hline 
		\multirow{2}{*}{Ce$_1$ (3/4, 3/4, 0.26424)} & $(0, 2, 0)$ & $(0, 0, 2)$                  \\
		& $+i(0, 0, 0)$                     & $+i(0, 0, 0)$                    \\ 
		\hspace{0.1cm}
		\multirow{2}{*}{Ce$_2$ (1/4, 1/4. 0.73576)} & $(0, 1.552, 0)$                    & $(0, 0, -1.552)$                    \\
		& $+i(0, 1.262, 0)$                     & $+i(0, 0, -1.262)$                     \\
		\hline 
		\hline
	\end{tabular}
	\label{tab:basisvectors}
\end{table}


\section{Discussion}

 From our magnetic susceptibility and neutron diffraction results, CeLiBi$_2$ must contain a $\Gamma_6$ CEF ground state Kramers doublet. To our knowledge, all other $\Gamma_6$ CEF ground state Ce$TX_2$ materials exhibit ferromagnetic ground states, and most of them contain Ce moments within the $ab$ plane. 

The notable exception to the latter pattern is CeAgSb$_2$ \cite{jang2019magnetic, takeuchi2003anisotropic, nikitin2021magnetic, prozorov2022topological}, wherein ferromagnetism arises at zero field from a $\Gamma_6$ ground state with Ce moments oriented along the $c$ axis \cite{jang2019magnetic, takeuchi2003anisotropic, nikitin2021magnetic, prozorov2022topological}.
Importantly, the low-energy spin dynamics of CeAgSb$_2$ measured by inelastic neutron scattering can be described within linear spin-wave theory by a combination of nine  anisotropic exchange interactions between local pseudo spin $S=1/2$ Ce moments within the $\Gamma_6$ ground state doublet \cite{nikitin2021magnetic}. 
These results point to the importance of long-range exchange interactions that go well beyond nearest neighbors.
More recently, a microscopic minimal Kondo-Heisenberg model uncovered the central role of long-ranged RKKY contributions in understanding magnetic order and magnon dispersion in AFM CeIn$_3$ \cite{cein3}.

In the case of CeLiBi$_2$, wherein Kondo or quantum critical effects are negligible, our results unravel the delicate balance between the crystal electric field ground state and multiple competing exchange interactions, which likely contain substantial long-range exchange contributions. In fact, long-range RKKY interactions in generic Fermi surfaces usually give rise to incommensurate magnetic structures \cite{cein3}, and we find evidence of an incommensurate AFM ordered state  in CeLiBi$_2$ contrasting with the $\Gamma_6$ CEF$-$FM order trend.

The best model of our neutron diffraction data is an elliptically modulated cycloidal structure where the Ce moments primarily reside within the $ab$ plane (Figure \ref{fig:scattering}). Ce moments are ferromagnetically coupled along the $a$ axis and antiferromagnetically modulated along $\bf{k}$ in the $b$ and $c$ axes. The size of the modulation within the magnetic unit cell varies from $1.24(6)$ $\mu_B$ Ce moments nearly parallel to the $ab$ plane to $0.58(3)$ $\mu_B$ Ce moments nearly parallel to the $c$ axis, producing a $\approx 0.66$ $\mu_B$ modulus magnitude. Because the Ce moments in CeLiBi$_2$ are highly localized with minimal Kondo coupling, the modulation of the Ce moment is not accomplished via itinerant $4f$ electrons near $E_\textrm{F}$ with a Fermi nesting vector in a spin density wave. Instead, the elliptical cycloid naturally forms as a consequence of the $\Gamma_6$ $g$-factor anisotropy and competing RKKY exchange interactions including short- and long-range components as discussed above. 

The magnetic peaks determined in Figure \ref{fig:scattering}(c) for CeLiBi$_2$ were described by a single incommensurate propagation wave vector along $ab$ plane as ${\bf{k}} = (0, 0.0724(4), 0.5)$ that produces an elliptical cycloidal magnetic structure. We note that neutron powder diffraction data are insensitive to the difference of multi-$\bf{k}$ magnetism (e.g., ${\bf{k_1}} = (0, 0.0724(4), 0.5)$ and  ${\bf{k_2}} = (0.0724(4), 0, 0.5)$) that could generate a modulated magnetic structure as seen in elemental Nd \cite{arachchige2022nanometric}. However, the single$-{\bf{k}}$ propagation wave vector is similar to those of CeAgBi$_2$ \cite{unpublished} and CeAuSb$_2$ \cite{marcus2018multi}, as AFM members of this family contain an ordering wave vector near ${\bf{k}} = (0, 0, 0.5)$ with incommensurability within the $ab$ plane. Notably, the modulation of these magnetic phases occurs in a length scale that is an order of magnitude larger than the lattice spacing, which again likely arises due to competitive long-range RKKY interactions. However, the zero-field propagation wave vector of CeLiBi$_2$ curiously coincides with the $b$ axis (or equivalently the $a$ axis), whereas in CeAgBi$_2$ and CeAuSb$_2$ the wave vector points between $a$ and $b$ as ${\bf{k}} = (0.18, 0.22, 0.5)$ and ${\bf{k}} = (0.136,\pm 0.136, 0.5)$, respectively. Similar wave vectors were attempted to model CeLiBi$_2$ but were unsuccessful. Interestingly, CeAuSb$_2$ under $b$ axis uniaxial pressure adopts a commensurate wave vector ${\bf{k}} = (0, 0.25, 0.5)$ where the in-plane component resides along the $b$ axis \cite{waite2022spin}. In this case, however, uniaxial stress explicitly breaks the underlying tetragonal symmetry of the lattice, which allows additional terms in the Landau free energy, but such symmetry breaking is not observed in CeLiBi$_2$. Future single crystal neutron scattering measurements could readily improve our understanding of this material by further refining the ordering wave vector with elastic measurements and investigate the balance between local CEF anisotropy and itinerant magnetic exchange in dynamic inelastic scattering experiments.

Another atypical characteristic of CeLiBi$_2$ is the hard-axis metamagnetic transition at $\mu_0H = 2$ T observed in magnetization and electrical resistivity data in Figure \ref{fig:FigureRT}. In general, a hard-axis transition is not a common property of $f$ electron materials, especially when observed at a lower field compared to metamagnetic transitions along the easy axis \cite{rai2018anomalous, aoki1998thermal}. For CeLiBi$_2$, two likely scenarios of hard-axis metamagnetism are a change in the moment size or magnetic structure. Magnetic moment size change could result from population of neighboring CEF states in a magnetic field, but the change in magnetization of CeLiBi$_2$ in Figure \ref{fig:Figure2}(d) is not significant enough to support this, which agrees with the well separated CEF ground state. 

If not a change in the CEF contribution, a natural scenario is that the multiple exchange interactions present in CeLiBi$_2$ are field dependent, and the out-of-plane magnetic field therefore alters the zero-field cycloidal magnetic structure. Magnetic field tuning of the competition between the single-ion CEF anisotropy and the anisotropy of exchange interactions has been recently proposed for YbRh$_3$Si$_7$, wherein metamagnetic transitions also occur for fields along the hard $c$ axis \cite{rai2018anomalous}. In addition, spin density wave materials such as Sr$_3$Ru$_2$O$_7$ \cite{lester2015field} and URu$_2$Si$_2$ \cite{knafo2016field} undergo wave vector changes in an applied field as the Fermi surface is modified. Magnetism in CeLiBi$_2$ is not a consequence of a spin density wave, but the magnetic structure is formed via extended RKKY exchange interactions that would be susceptible to Fermi surface changes. Alternatively, the hard-axis magnetic field may push CeLiBi$_2$ towards a commensurate 'lock-in' phase transition as seen in CeAuSb$_2$ under pressure \cite{waite2022spin}, in spin density wave materials like CaFe$_4$As$_3$ as function of temperature \cite{manuel2010incommensurate}, and CeRh$_3$Si$_2$ \cite{amorese2022metamagnetism} or other incommensurate spiral magnetic materials like LiYbO$_2$ \cite{bordelon2021frustrated} in a magnetic field. Finally, a change in the spin texture is a related viable explanation of this hard-axis transition in CeLiBi$_2$ akin to other Ce$TX_2$ materials as a function of temperature, pressure, or field. For example, magnetic skyrmion formation has been recently observed in CeAgBi$_2$ in an external magnetic field above $\mu_0H = 4$ T in polarized neutron scattering experiments \cite{unpublished}. Possible development of a multi$-\bf{k}$ structure in a magnetic field in CeLiBi$_2$ prompts further single crystal neutron scattering measurements in a magnetic field to fully determine the origin of this metamagnetic transition. 

\begin{figure}[!t]
	\includegraphics[width=3in]{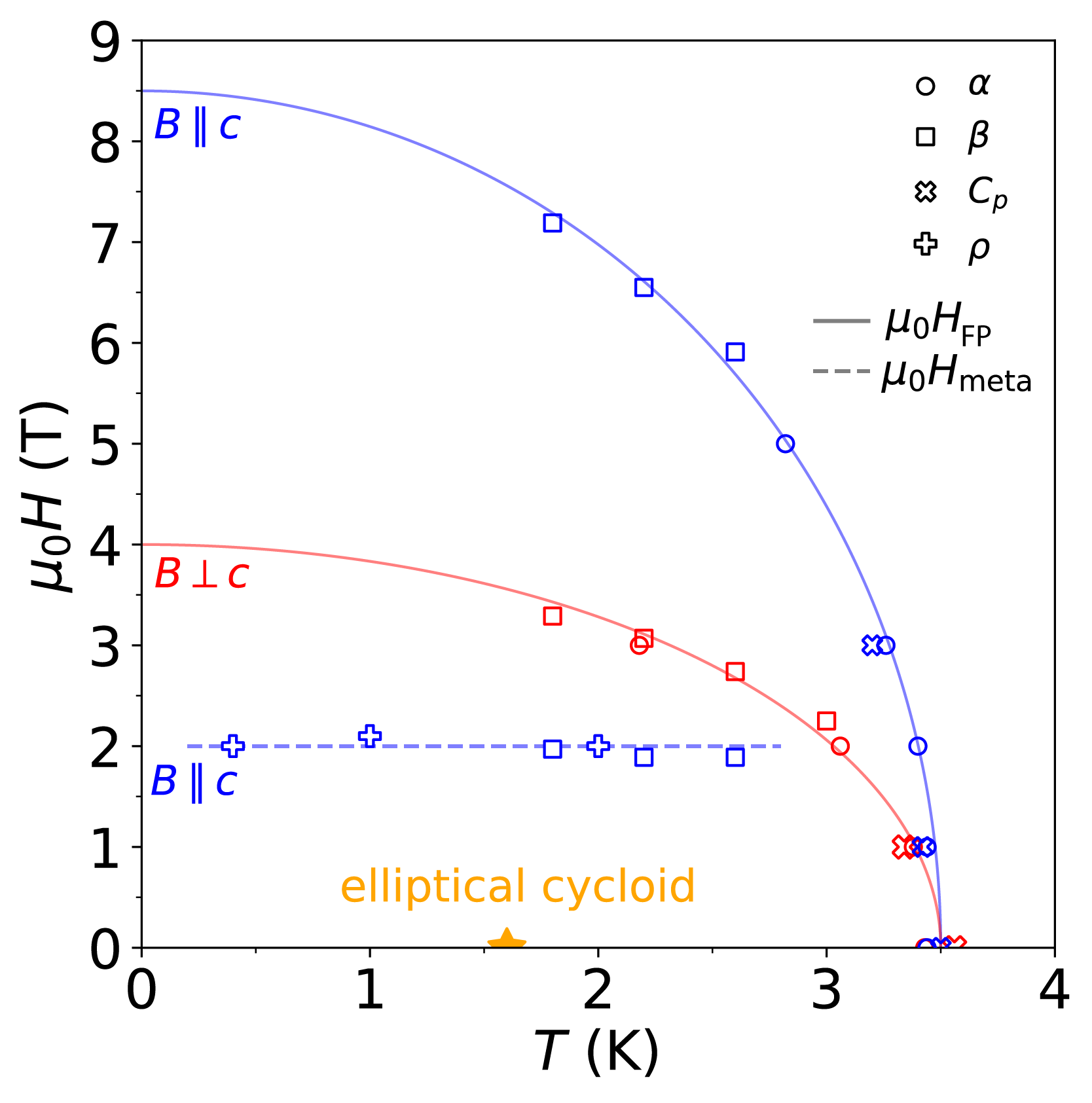}
	\caption{Phase diagram of CeLiBi$_2$ obtained by combining thermal expansion ($\alpha$), magnetostriction ($\beta$), specific heat ($C_p$), and electrical resistivity ($\rho$) measurements. Blue (red) curves represent data collected in magnetic fields parallel (perpendicular) to the $c$ axis. Solid lines represent the phase boundaries between the antiferromagnetically ordered elliptical cycloid phase to the field polarized regime. The dashed blue line indicates the location of the hard-axis metamagnetic transition. The golden star indicates where elastic neutron powder diffraction data determined the elliptical cycloid magnetic structure in CeLiBi$_2$.}
	\label{fig:phasediag}
\end{figure}

The hard-axis metamagnetic transition also appears in resistivity and magnetostriction measurements in Figures \ref{fig:FigureRT}(e)-(f) and \ref{fig:FigureTE}(c) and is summarized in Figure \ref{fig:phasediag}. These ubiquitous signatures suggest that the electronic and magnetic properties of CeLiBi$_2$ are moderately, but undoubtedly, intertwined. CeLiBi$_2$ resides within the high-conductivity regime ($\sigma_{0} \sim 0.71 \times 10^6$~($\Omega$~cm)$^{-1}$), which suggests skew scattering is the dominant process in this clean material. Skew scattering is in stark contrast to magnetic scattering, which would cause a reduction in the resistivity of CeLiBi$_2$ as a function of magnetic field due to increased spin polarization. Instead, we observe a large positive MR $\sim 650\%$ at $\mu_0H = 9$~T (Figure \ref{fig:FigureRT}(c)-(d)). Nonsaturating MR can arise, for instance, in a nearly compensated semimetal with quadratic (linear) band dispersion producing quadratic (linear) MR field dependence \cite{pippard1989magnetoresistance, leahy2018nonsaturating}. This appears in semimetals such as (Ta,Nb)(Sb,As)$_2$ and (Ta,Nb)P \cite{leahy2018nonsaturating, taasnbas}. Alternatively, nonsaturing linear MR appears in clean metals displaying quantum oscillations in the limit $\omega_c \tau >> 1$  \cite{nagaosa2010anomalous, PhysRevB582788, abrikosov2000quantum} as is found in the high-field regime of CeCoIn$_5$ \cite{PhysRevLett91246405, PAGLIONE2004705}. CeLiBi$_2$ is a clean metal with light carriers in the high-conductivity limit and therefore the condition $\omega_c \tau >> 1$ is likely to be satisfied.
 
We determined the mass of the carriers in CeLiBi$_2$ via magnetostriction and dHvA measurements displaying QO up to $T = 30$ K and $\mu_0H = 55$ T. Further, the frequency of the slowest oscillations and associated effective mass of $m^* \approx 0.07 \ m_e$ agree with results from other materials of the same structure type such as (Ba,Ca,Sr)Mn(Sb,Bi)$_2$, LaAgBi$_2$, and other Ce$TX_2$ materials \cite{park2011anisotropic, lee2013anisotropic, wang2012two, wang2013quasi, farhan2014aemnsb2, he2017quasi, liu2016nearly, jeong2006electronic, alsardia2020pressure}. In the absence of spin-orbit coupling, the square nets of Bi ions form Dirac cones with $p$ orbital character near $E_\textrm{F}$. As has been noted previously in SrMnBi$_2$, the presence of strong spin-orbit coupling from the heavy Bi atoms can create gaps at the Dirac points of around 40 meV \cite{park2011anisotropic}. These band crossings become avoided band crossings, generating small electron pockets near $E_\textrm{F}$ near ($\pi$,0,0) and along (0,0,$\pi$) to ($\pi$,0,$\pi$) in Figure \ref{fig:DFT}. Consequently, these electron pockets are not expected to contain Dirac fermions, but their highly dispersive bands can contain small effective mass carriers that, depending on the exact tuning of $E_\textrm{F}$, are the most likely origin of the light carriers in CeLiBi$_2$. The coexistance of light, mobile carriers based on Bi square nets and AFM order makes CeLiBi$_2$ a possible platform for developing spintronic devices\cite{bhatti2017spintronics} as has been proposed for GdTe$_3$ containing an analogous Te square net and AFM order \cite{doi:10.1126/sciadv.aay6407}. To explore this further, angle-resolved photoemission spectroscopy and angle-resolved dHvA measurements are desired.

\section{Conclusions}

CeLiBi$_2$ expands the Ce$TX_2$ family into the alkali metal phase space and opens a new chemical pathway to tune the physical properties in this family. Our combined single crystal physical properties measurements and elastic neutron powder diffraction analyses show that CeLiBi$_2$ hosts a breadth of atypical magnetic and electronic properties. First, contradicting $\Gamma_6$ CEF and antiferromagnetic incommensurate cycloidal order below $T_N = 3.4$ K suggests a delicate balance between CEF anisotropy and competing long-ranged magnetic exchange interactions. Second, a hard-axis metamagnetic transition occurs near $\mu_0H = 2$ T in magnetization, magnetostriction, and transport measurements implying intertwined changes in magnetic structure or texture with electronic properties. Third, CeLiBi$_2$ is a rare example of a clean material with high conductance, a large anomalous Hall effect, and positive linear magnetoresistance, indicating that skew scattering dominates in the material. Fourth, we find evidence of quantum oscillations at five different frequencies arising from light carriers from square-net Bi bands. All of these properties separately are uncommon, but not unique. However, their coexistence places newly-synthesized CeLiBi$_2$ as a prime candidate for investigating their complex interplay in a single clean material.

\begin{acknowledgments}
	We would like to thank M. Janoschek, D. Yahne, W. Simeth, and C. Batista for fruitful discussions. Work at Los Alamos was performed	under the auspices of the U.S. Department of Energy, Office of Basic Energy Sciences, Division of Materials Science and Engineering. MB, CG, and SMT acknowledge support from the Laboratory
	Directed Research and Development program. Scanning electron microscope and energy dispersive X-ray measurements were performed at the Center for Integrated Nanotechnologies, an Office of Science User Facility operated for the U.S. Department of Energy Office of Science. Pulsed field measurements (NH and RK) were supported by the Department of Energy (DoE) BES project `Science of 100 tesla.' The National High Magnetic Field Laboratory, which hosts the high magnetic field magnets, is funded by NSF Cooperative Agreements DMR-1157490 and 1164477, the State of Florida and DoE. A portion of this research used resources at the High Flux Isotope Reactor a DOE Office of Science User Facility operated by the Oak Ridge National Laboratory.

\end{acknowledgments}

\bibliography{CLB_bib_v4_abbr}

\end{document}